\documentclass[journal]{IEEEtran}
\usepackage[numbers,sort&compress]{natbib}
\usepackage{amsmath,amssymb,amsfonts}
\usepackage{graphicx}
\usepackage{subfigure}
\usepackage[section]{placeins}
\usepackage{float}
\usepackage{multirow}
\usepackage{textcomp}
\usepackage{algorithm}
\usepackage{algpseudocode}
\usepackage{amsmath}
\usepackage{ntheorem}
\usepackage{graphics}
\usepackage{epsfig}
\usepackage{threeparttable}
\usepackage{amsmath}
\usepackage{bm}
\usepackage{tabularx}
\usepackage{multirow}
\usepackage{threeparttable}
\usepackage{mathrsfs}
\usepackage{array}
\usepackage{color}
\usepackage{epstopdf}

\ifCLASSINFOpdf
\else
\fi
\begin{document}
\title{Dual-level Semantic Transfer Deep Hashing\\ for Efficient Social Image Retrieval}
\author{Lei Zhu,
        Hui Cui,
        Zhiyong Cheng,
        Jingjing Li,
        Zheng Zhang
\thanks{L. Zhu and H. Cui are with the School of Information Science and Engineering, Shandong Normal University, Jinan 250358, China}
\thanks{Z. Cheng is with Shandong Computer Science Center (National Supercomputer Center in Jinan), Qilu University of Technology (Shandong Academy of Sciences)}
\thanks{J. Li is with the School of Computer Science and Engineering, University of Electronic Science and Technology of China, Chengdu 611731, China.}
\thanks{Z. Zhang is with Bio-Computing Research Center, Harbin Institute of Technology, Shenzhen 518055, China.} 
\thanks{Copyright©2020 IEEE. Personal use of this material is permitted. However, permission to use this material for any other purposes must be obtained from the IEEE by sending an email to pubs-permissions@ieee.org.}
}

\markboth{IEEE Transactions on Circuits and Systems for Video Technology}%
{Shell \MakeLowercase{\textit{et al.}}: Bare Demo of IEEEtran.cls for IEEE Journals}

\maketitle

\begin{abstract}
Social network stores and disseminates a tremendous amount of user shared images.
Deep hashing is an efficient indexing technique to support large-scale social image retrieval,
due to its deep representation capability, fast retrieval speed and low storage cost.
Particularly, unsupervised deep hashing has well scalability as it does not require any manually labelled data for training.
However, owing to the lacking of label guidance,
existing methods suffer from severe semantic shortage
when optimizing a large amount of deep neural network parameters.
Differently, in this paper, we propose a \emph{Dual-level Semantic Transfer Deep Hashing} (DSTDH) method to
alleviate this problem with a unified deep hash learning framework.
Our model targets at learning the semantically enhanced deep hash codes by specially exploiting the
user-generated tags associated with the social images.
Specifically, we design a complementary dual-level semantic transfer mechanism
to efficiently discover the potential semantics of tags and seamlessly transfer them into binary hash codes.
On the one hand, instance-level semantics are directly preserved into hash codes from the associated tags with adverse noise removing.
Besides, an image-concept hypergraph is constructed for indirectly transferring the latent high-order semantic correlations of images and tags into hash codes.
Moreover, the hash codes are obtained simultaneously with the deep representation learning
by the discrete hash optimization strategy.
Extensive experiments on two public social image retrieval datasets validate the superior performance of our method compared with state-of-the-art hashing methods.
The source codes of our method can be obtained at https://github.com/research2020-1/DSTDH
\end{abstract}

\IEEEpeerreviewmaketitle

\section{Introduction}

\IEEEPARstart{N}{owadays}, millions of individuals share their daily photos in the social networks.
This trend leads to an explosive increase of social images.
It is important to develop advanced indexing techniques to support efficient retrieval from such large-scale social image databases.

Benefited by recent representation learning \cite{CrossMod1,SemiSuper1,CrossMod2,TransferHash1,SemanticHash1,CrossMod3,CrossMod4,review21,review22,review23,review24,Hong1,Hong2},
hashing becomes popular in large-scale image retrieval.
The hashing technique encodes the high-dimensional images into low-dimensional binary codes.
It has the advantages of fast Hamming distance calculation and low storage cost,
so it can greatly speed up the large-scale image retrieval.
Generally, existing hashing approaches can be divided into two families:
supervised and unsupervised hashing \cite{CrossMod5,KSH,FastH,SDH,SH,ITQ,SGH}.
Supervised hashing \cite{KSH,FastH,SDH} learns hash codes by explicit semantic labels,
which can strengthen the discriminative semantics of hash codes.
However, the expensive annotation cost of semantic labels significantly reduces its scalability.
Unsupervised hashing \cite{SH,ITQ,SGH} only exploits the feature information of data
without relying on any supervised labels.
Compared with supervised hashing, it has more desirable scalability,
because most images in real-world large-scale retrieval scenarios may have no or limited manually annotated labels.
Nevertheless, without the supervision of the explicit semantic labels,
the unsupervised hash codes could be subject to the limited semantics,
which reduces the hash performance.
It is a challenge that how to learn discriminative binary hash codes while maintaining the well application scalability (reducing the label dependance).
\begin{figure*}
\centering
\subfigure{\includegraphics[scale=0.8]{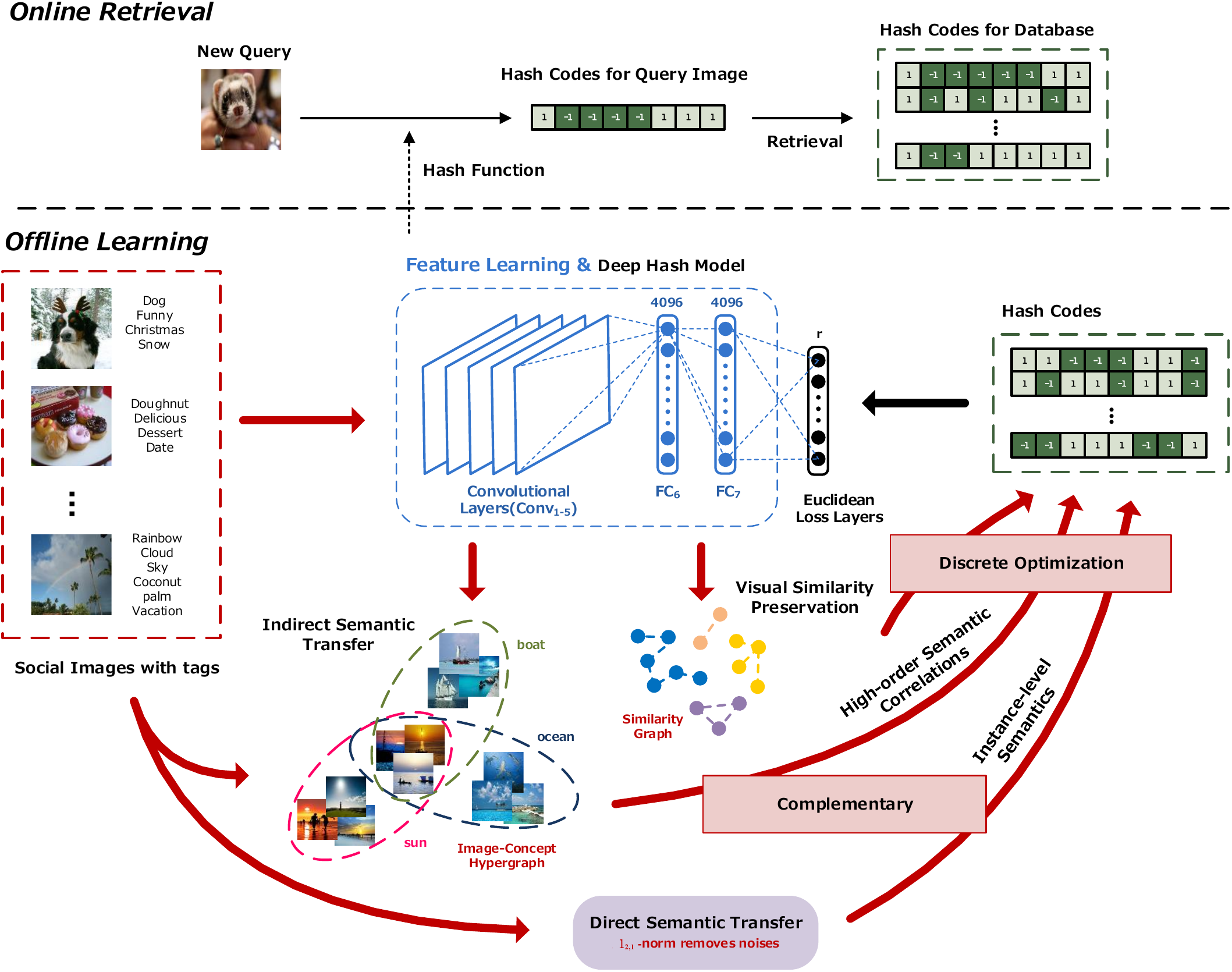}}
\caption{The basic framework of dual-level semantic transfer deep hashing method. The retrieval system consists of two parts. The offline part trains a deep hash model,
which can both generate image feature representations and project images into hash codes.
The features of images assist in updating the image-concept hypergraph and the similarity graph.
The semantics of hash codes can be strengthened by image-concept hypergraph, similarity graph, and denoised tags.
Besides, the hash codes are directly learned based on the efficient discrete optimization with low computation and storage cost.
The online part first obtains the hash codes of a new query image by the hash functions learned at offline stage.
Then, the Hamming distances between the hash codes of the query image and those of database samples are calculated.
Finally, the database samples are sorted in the ascending order of their corresponding Hamming distances.
}
\label{Figfram}
\end{figure*}

Recent literature witness the birth and success of unsupervised deep hashing \cite{DeepBit,UH-BDNN,SADH,GreedyHash},
which enhances the representation capability of hash codes via joint deep image representation and hash learning.
Nevertheless, existing unsupervised deep hashing methods still suffer from various shortcomings.
Several works train the deep neural network by directly minimizing the quantization loss \cite{DeepBit}
or data reconstruction errors \cite{UH-BDNN}.
They fail to preserve the similarities of images into the hash codes.
\cite{SADH} learns the hash codes by considering the visual relations in deep neural network framework. But,
the semantics of the learned hash codes are still insufficient.
Besides, it explicitly computes the graph for visual modeling, which is time-consuming when optimizing the hash codes.
\cite{GreedyHash} employs the greedy principle to solve deep hash optimization problem,
which is also time-consuming.

Fortunately, social images are generally accompanied with informative tags voluntarily provided by users \cite{WMH}.
These freely provided tags are semantically relevant to the image contents.
Technically, they provide rich free semantic sources to enhance the discriminative capability of hash codes.
With this motivation, several hashing works \cite{SaH,WMH,weakly3} have been specifically developed for social image retrieval.
\cite{SaH} and \cite{WMH} are shallow hashing methods. They simply learn the hash codes by hand-crafted features and social tags.
However, the hand-crafted features involve limited semantics of images.
In addition, these two methods directly exploit the social tags for hash learning.
They ignore the involved adverse noises that may negatively affect the hashing performance.
\cite{weakly3} proposes a deep hashing framework consisting of the weakly-supervised
pre-training stage and the supervised fine-tuning stage.
Because it uses label information to assist hash code learning,
it has the same label dependence as supervised hashing.

Discrete optimization is important for generating hash codes.
It is very hard to handle the hash optimization problem because of the binary constraints on hash codes.
Two common optimization strategies are adopted in existing methods.
The first is two-step relaxing+rounding optimization strategy
(social hashing methods \cite{SaH}, \cite{WMH} and \cite{weakly3} adopt this approach).
It relaxes the discrete constraints on hash codes firstly, and then solves the finally hash codes by mean-thresholding.
This hash optimization method may produce significant quantization errors,
which will lead to suboptimal solutions.
The other is the bit-by-bit hash optimization
based on Discrete Cyclic Coordinate Descent (DCC) \cite{SDH}.
It is still time consuming because each step only optimizes one bit and multiple iterations are required to learn all hash bits.

Motivated by above analyses, in this paper,
we propose a \emph{Dual-level Semantic Transfer Deep Hashing} (DSTDH) method
for efficient social image retrieval.
The highlights of this paper are summarized below:

\begin{itemize}
  \item We develop a scalable weakly-supervised deep hash learning framework that targets at learning semantically enhanced hash codes via complementary dual-level semantic transfer. Instance-level semantics are discovered from the associated social tags with adverse noise removal for direct semantic transfer. Besides, an image-concept hypergraph is constructed to indirectly transfer the complementary high-order semantic correlations of images and tags into the hash codes. As far as we know, there is still no similar work.\vspace{1mm}
  \item To guarantee the transfer effectiveness, an Augmented Lagrangian Multiplier (ALM) \cite{ALM} based binary optimization method is proposed to tackle the formulated hash learning problem.
      The hash codes are directly and efficiently solved to alleviate the relaxing quantization loss by a simple operator rather than the iterative hash code-solving step.\vspace{1mm}
  \item Experiments on two public social image retrieval benchmarks demonstrate the superiority of our method. Extensive ablation analysis validate the advantages of our model from various aspects.
\end{itemize}

Note that, our method is different from
multi-modal fused hashing \cite{multimodal1,multimodal2,multimodal3,multimodal4,ZHU1,ZHU2,ZHU3}
and cross-modal hashing methods \cite{review31,review32,review33}
that exploits heterogeneous modalities to learn compact hash codes.
Multi-modal fused hashing methods collaborate the multi-modal features of the samples
at both training and query stages to learn the hash codes.
Cross-modal hashing methods are designed for
the large-scale cross-modal retrieval scenarios \cite{review41,review42}
where the query and its retrieval results are from different modality types.
In contrast, our method mainly aims at learning the image hash codes
with the weak supervision from the accompanied social tags,
the training setting and learning objective is different from them.
Also, our method is different from domain transfer learning methods \cite{review1,Transfer5,Transfer1,Transfer2,Transfer3,Transfer4}
that try to acquire knowledge from the source domain to improve the learning performance in the target domain.
The source domain and target domain usually come from different resources
which may have the same or different data distribution.
Contrastively, in our paper, we devote to learning compact image hash codes
with the weak supervision from the accompanied social tags.
There is no hash learning task on the source domain (social tags). Thus, the learning process is different from the general domain transfer learning.

\section{Related Work}\label{relwork}
\subsection{Supervised Hashing}
The supervised hashing methods take advantage of the discriminative semantic labels as supervised information
to learn the hash codes.
Representation shallow hashing methods include:
Iterative Quantization with Canonical Correlation Analysis (CCA-ITQ) \cite{ITQ},
Supervised Discrete Hashing (SDH) \cite{SDH},
Supervised Hashing with Kernels (KSH) \cite{KSH},
and Latent Factor Hashing (LFH) \cite{LFH}.
Typical deep hashing methods include:
Convolutional Neural Network Hashing (CNNH) \cite{CNNH},
Network In Network Hashing (NINH) \cite{NINH},
Deep Pairwise Supervised Hashing (DPSH) \cite{DPSH},
Deep Supervised Discrete Hashing (DSDH) \cite{DSDH},
Supervised Semantics-preserving Deep Hashing (SSDH) \cite{SSDH} and
Supervised Discrete Hashing with Relaxation (SDHR) \cite{SDHR}.
Under the supervision of explicit semantic labels,
the performance of the supervised hashing methods is superior to the unsupervised approaches.
Nevertheless, the heavy burden of manually annotating semantic labels
hinders the application of supervised hashing methods in large-scale image retrieval.

\subsection{Unsupervised Hashing}
Unsupervised hashing does not utilize semantic labels to learn hash functions during the training process.
It tries to preserve visual data structure into hash codes.
Spectral Hashing (SH) \cite{SH} first learns the principal directions of the samples, and then solves the hash codes via spectral analysis.
Principal Component Analysis Hashing (PCAH) \cite{PCAH} directly adopts the eigenvectors of principal component of the data as the projection vectors to solve the hash codes.
Iterative Quantization (ITQ) \cite{ITQ} first learns the projected value by principal component analysis,
and then transforms them into hash codes by finding a orthogonal rotation matrix.
Scalable Graph Hashing (SGH) \cite{SGH} does not explicitly calculate the similarity graph matrix. It uses a sequential learning method to learn the hash codes in a bit-wise manner.
Latent Semantic Minimal Hashing (LSMH) \cite{LSMH} first learns latent semantic feature based on matrix decomposition,
and then explores the refined latent semantic features embedding to
generate hash codes

Unsupervised deep hashing methods show promising performance improvement compared to the shallow methods
with the joint image representation learning and hash coding.
\cite{UH-BDNN} designs an Unsupervised Hashing with Binary Deep Neural Network (UH-BDNN)
which discards the binary constraints and
learns hash codes as the outputs of a binary auto-encoder.
In \cite{DeepBit}, DeepBit is proposed to enforce three losses at the last layer of the neural network
to learn the binary descriptors. Despite the success, these two pioneering methods fail to explicitly keep the similarities among data
in the learned hash codes, so they are far from satisfactory performance (as shown in experiments).
In Similarity-Adaptive Deep Hashing (SADH) \cite{SADH}, the hash codes are learned by iteratively constructing similarity graph and performing hash optimization.
It progressively improves the hashing performance.
Nevertheless, as only considering the visual information for image retrieval,
SADH still suffers from insufficient semantics.
Moreover, it explicitly computes the visual graph, which is time-consuming during the optimization process.
In GreedyHash \citep{GreedyHash}, the cosine distances between data samples are minimized in different space to preserve the data similarity.
It uses a time-consuming greedy principle to solve the hash optimization problem.

\subsection{Social Image Hashing}
There are few hashing methods specially designed for social image retrieval.
Semantic-Aware Hashing (SAH) \cite{SaH}
employs heterogeneous information such as textual (tag) and visual (image) domains
to obtain semantic enhanced binary hash codes.
Weakly-supervised Multi-modal Hashing (WMH) \cite{WMH}
learns hash functions by exploring tag information, local discriminative information
and geometric structure in visual space simultaneously.
Both SAH and WMH strengthen the discriminative semantics of hash codes by exploring the freely available social tags.
However, in these methods,
the image feature extraction process and hash function learning process are performed separately. Thus, the features may be not fully compatible with the hash learning.
In addition, the linear hash functions they adopted
may limit the representation capability of the learned hash codes.
Moreover, they fail to alleviate the noises from social tags that may negatively affect the hashing performance.
Finally, they adopt relaxed hash optimization method that may produce significant quantization errors to learn hash codes.
\cite{weakly3} proposes a deep hash framework for social image retrieval.
It is mainly made up of two phases:
weakly-supervised pre-training process and supervised fine-tuning process.
In this framework, the semantic embedding vector is introduced for each image.
The hashing learning and the semantic vector learning are performed jointly.
The performance of this method is also dependent on supervised labels, which limits its scalability.

Differently, we propose a social image hashing method in this paper to generate discriminative binary hash codes
by simultaneously considering deep learning framework, the denoised user-generated social tags,
and complementary dual-level semantic transfer.

\section{Our Method}
\subsection{Notations Definition}
Throughout the paper,
the vectors are represented as boldface lowercase characters and
the matrices are represented as boldface uppercase characters.
For matrix $\mathbf{A} \in \mathbb{R}^{m \times n}$, the ${i}_{th}$ column and the $\left( {i,j} \right)_{th}$ element are represented by
${\mathbf{a}_i}$ and $a_{ij}$, respectively.
\texttt{Tr}($\mathbf{A}$) denotes the trace of $\mathbf{A}$.
$\mathbf{A}^\texttt{T}$ is the transpose of $\mathbf{A}$.
The Frobenius norm and the $\ell_{2,1}$-norm of $\mathbf{A}$ are denoted by
${\lVert \mathbf{A} \rVert_\texttt{F}^2} = \sum\nolimits_{i = 1}^m {\sum\nolimits_{j = 1}^n {a_{ij}^2}}$ and
${\lVert \mathbf{A} \rVert_{2,1}} = \sum\nolimits_{i = 1}^m {\sqrt {\sum\nolimits_{j = 1}^n {a_{ij}^2}}}$, respectively.
The diagonal matrix operator is denoted as $\texttt{diag}(\cdot)$.
The sign function is represented by $\texttt{sgn}(\cdot)$,
which returns $1$ for positive, and $-1$ otherwise.
$\mathbf{I}_n \in \mathbb{R}^{n \times n} $ is the identity matrix, and
$\mathbf{1}$  denotes the vector in which all the elements are 1.

Suppose that $\bm{{\rm X}} = \left[ {\bm{{\rm x}}_1,\cdot \cdot \cdot ,\bm{{\rm x}}_n} \right] \in {\mathbb{R}^{d \times n}}$ is a social image dataset,
which contains \emph{n} images.
Each image ${\bm{{\rm x}}_i \in \mathbb{R}^d}$ (\emph{d} is the dimensionality of image feature)
is assigned with \emph{c} non-overlapping social tags by users.
We denote the image-tag relation as $\bm{{\rm Y}} = \left[ {\bm{{\rm y}}_1, \cdot  \cdot  \cdot, \bm{{\rm y}}_n} \right] \in {\mathbb{R}^{c \times n}}$.
The tag set of each image as $\mathbf{y}_i\in \mathbb{R}^c$,
where $y_{ji}= 1$ if $\bm{{\rm x}}_i$ is associated with ${j}$th tag and $y_{ji}= 0$ otherwise.
Our objective is to learn the hash function $h(\cdot)$,
which can generate ${r}$-bits hash codes $\bm{{\rm Z}} = \left[ {\bm{{\rm z}}_1, \cdot  \cdot  \cdot, \bm{{\rm z}}_n}  \right] \in {\left[ { - 1, 1} \right]^{r \times n}}$
for social images.
The main notations and system overview are shown in Table \ref{table1} and Fig.\ref{Figfram}, respectively.

\subsection{The Proposed Model} \label{sub3_3}
\subsubsection{Visual Similarity Preservation}
\begin{table}
\caption{Summary of main notations}
\centering
\setlength{\tabcolsep}{1.5mm}{
\begin{tabular}{l|l}
\hline
\textbf{Notation} & \textbf{Explanation} \\
\hline
\emph{n}       & number of images                 \\
\emph{r}       & hash code length                 \\
\emph{m}       & number of anchors                \\
\emph{a}       & number of concepts               \\
$\mathbf{X}$   & visual feature matrix of images  \\
$\mathbf{Y}$   & image-tag relation matrix        \\
$\mathbf{Z}$   & hash codes matrix                \\
$\mathbf{P}$   & semantic transfer matrix         \\
$\mathbf{L}_G$ & visual graph Laplacian matrix \\
$\mathbf{L}_H$ & hypergraph Laplacian matrix  \\
$\mathbf{A}$, $\mathbf{B}$, $\mathbf{E}_A$, $\mathbf{E}_B$ &  auxiliary matrices of ALM \\
$\mathbf{H}$   & incidence matrix of hypergraph                  \\
$\mathbf{D}_v$ & vertex degree matrix in hypergraph \\
$\mathbf{D}_e$ & edge degree matrix in hypergraph   \\
$\mathbf{D}_w$ & edge weight matrix in hypergraph   \\
\hline
\end{tabular}
}
\label{table1}
\end{table}
The objective of social image retrieval is to retrieve semantically similar images for the query. Hence, mapping visually similar social images to the binary codes with the short Hamming distance is significant for the retrieval performance.
In this paper, we resort to graph model \cite{SGH,graph1,graph2} to preserve
semantic similarities of images in binary hash codes.
Specifically, heavy penalty will be incurred if two similar images are mapped far apart.
To this end, we seek to minimize the weighted Hamming distance of hash codes
\begin{equation}\label{eq32}
\begin{aligned}
&\mathop {\min }\limits_{\mathbf{Z} \in {\left\{ { -1,1} \right\}^{r \times n}}} \sum\limits_{i,j = 1}^n {} {{s_{ij}}\left\| {{{\bf{z}}_i} - {{\bf{z}}_j}} \right\|_\texttt{F}^2}, \\
 \Leftrightarrow &\mathop {\min }\limits_{\mathbf{Z} \in {\left\{ { -1,1} \right\}^{r \times n}}} \texttt{Tr}({\bf{Z}}{{\bf{L}}_G}{{\bf{Z}}^\texttt{T}}) ,
\end{aligned}
\end{equation}
where $\mathbf{S} \in \mathbb{R}^{n \times n} $
is the affinity matrix of  visual graph,
$\mathbf{L}_G = \texttt{diag}(\mathbf{S1}) - \mathbf{S}$ is the corresponding Laplacian matrix.

The calculation of $\mathbf{S}$ and $\mathbf{L}_G$ will consume $O(n^2)$,
which is unacceptable in large-scale social image retrieval.
In this paper, we adopt anchor graph \cite{AGH} to avoid this problem.
Specifically, we approximate the affinity matrix as $\mathbf{S} = \mathbf{V}\mathbf{\Lambda}^{-1} {\mathbf{V}^{\texttt{T}}}$,
where $\mathbf{V} \in {\mathbb{R}^{n \times m}}$ is a matrix which is obtained by computing the similarities between images and \emph{m} anchor points,
and $\mathbf{\Lambda} = \texttt{diag}({\mathbf{V}^{\texttt{T}}}\mathbf{1})\in {\mathbb{R}^{m \times m}}$.
The resulted graph Laplacian matrix can be represented as
$\mathbf{L}_G = \mathbf{I}_n - \mathbf{V}{\mathbf{\Lambda}^{-1}}{\mathbf{V}^{\texttt{T}}}$.
According to above analysis,
we rewrite Eq.(\ref{eq32}) as the following equivalent form:
\begin{equation}
\begin{aligned}
&\mathop {\min_{\mathbf{Z} \in {\left\{ { -1,1} \right\}^{r \times n}}}}\texttt{Tr}(\mathbf{Z}(\mathbf{I}_n - \mathbf{V}\mathbf{\Lambda} {\mathbf{V}^{\texttt{T}}}){\mathbf{Z}^{\texttt{T}}}) \\
\Rightarrow &\mathop {\min_{\mathbf{Z} \in {\left\{ { -1,1} \right\}^{r \times n}}}}
-\texttt{Tr}(\mathbf{Z}\mathbf{V}\mathbf{\Lambda}^{-1} {\mathbf{V}^{\texttt{T}}}{\mathbf{Z}^{\texttt{T}}}).
\end{aligned}
\label{eq33}
\end{equation}

\subsubsection{Direct Semantic Transfer}
Due to the semantic gap,
visual feature inherently has the limitation on representing high-level semantics.
As aforementioned above, social images are usually associated with user-provided tags,
which generally contain high-level semantics from users.
Therefore, in this paper, we exploit the auxiliary social tags
and directly transfer their semantics into hash codes.
Specifically, we directly correlate the hash codes with tags with a semantic transfer matrix $\mathbf{P}$,
in order to directly transfer the involved instance-level semantics into the hash codes.
$\mathbf{P}$ is defined as $\mathbf{P} = \left[ {\mathbf{p}_1, \cdot  \cdot  \cdot, \mathbf{p}_r} \right] \in {\mathbb{R}^{c \times r}}$,
where $\mathbf{p}_i \in {\mathbb{R}^{c \times 1}}$ is the semantic transfer vector for the \emph{i}-bits hash codes.
Formally, we optimize the following problem to learn $\mathbf{P}$
\begin{equation}
\mathop {\min }\limits_{\mathbf{P}} \  \lVert {\mathbf{Z} - {\mathbf{P}^{\texttt{T}}}\mathbf{Y}} \rVert_{2,1}.
\label{eq11}
\end{equation}
Eq.(\ref{eq11}) adopts $\ell _{2,1}$-norm \cite{L21},
which is designed to automatically remove the noises from social tags.
The tags with stronger semantic discriminative capability are retained for semantic transfer.

\subsubsection{Indirect Semantic Transfer}
Social images are stored and disseminated in social networks.
The latent semantic correlations of images are important for social image retrieval.
However, the above visual graph and direct semantic transfer fail to fully capture the semantic correlations,
due to the well-known semantic gap and instance-level semantic transfer.
Hence, it is promising to involve the semantic correlations of images characterized by social tags into the hash codes.
Intrinsically, the latent semantic correlations of images are high-order.
It is common that a single social image will be described by multiple tags,
and a tag may be shared by multiple social images (as shown in Fig.\ref{Figfram}).
Theoretically, social images that share more tags
are more likely to have similar visual contents.

Inspired by this observation, in this paper,
we propose to construct an image-concept hypergraph for indirectly transferring the semantic correlations of images.
We first pack the visual feature matrix with the image-tag relation matrix,
and then detect $a$ concepts by $k$-means \cite{kmeans} on the integrated matrix.
In this case, the images are transformed as
$\hat{\mathbf{X}}= \left[ {\hat{\mathbf{x}}_1,\cdot \cdot \cdot ,\hat{\mathbf{x}}_n} \right] \in {\mathbb{R}^{(d+c) \times n}}$.
The detected concepts are represented as
$\mathbf{E} = \left[ {\mathbf{e}_1,\cdot \cdot \cdot ,\mathbf{e}_a} \right] \in {\mathbb{R}^{(d+c) \times a}}$.
To model the latent high-order semantic correlations of images,
we determine the images as vertices, and the concepts as hyperedges.
They jointly comprise the image-concept hypergraph.
In this way, an image represents several semantic concepts,
and several images may be latently included with more than one concept.
Under such circumstance,
the high-order semantic correlations are naturally modelled.
Mathematically,
the hypergraph can be represented with a $\emph{n} \times \emph{a}$ incidence matrix $\mathbf{H}$.
The incidence value between vertex ${\hat{\mathbf{x}}}\in {\hat{\mathbf{X}} } $ and hyperedge ${\mathbf{e}}\in {\mathbf{E}}$ in $\mathbf{H}$ is calculated as
\begin{equation}
h({\hat{\mathbf{x}}},{\mathbf{e}}) = \texttt{exp}( - \left\| {{\hat{\mathbf{x}}} - {\mathbf{e}}} \right\|_\texttt{F}^2/{2\sigma ^2}),
\end{equation}
where $\sigma$ is bandwidth parameter.
With $\mathbf{H}$, the degree of hyperedge ${\mathbf{e}}$ is calculated as
\begin{equation}
\delta (\mathbf{e}) =\sum\limits_{{\hat{\mathbf{x}}}\in {\hat{\mathbf{X}}}} h({\hat{\mathbf{x}}},{\mathbf{e}}).
\end{equation}
We assume that the concepts are evenly distributed in database.
Thus, we set the weights of all hyperedges to 1, $w({\mathbf{e}})=1$.
Accordingly, the degree of vertex ${\hat{\mathbf{x}}}$  is calculated as
\begin{equation}
d (\mathbf{\hat{\mathbf{x}}}) =\sum\limits_{{\mathbf{e}}\in {\mathbf{E}}} w({\mathbf{e}}){h({\hat{\mathbf{x}}},{\mathbf{e}})}
=\sum\limits_{{\mathbf{e}}\in {\mathbf{E}}}{h({\hat{\mathbf{x}}},{\mathbf{e}})}.
\end{equation}

Principally, images that belong to more identical hyperedges
will describe similar semantic concepts with greater probability.
Therefore, they should be mapped to similar binary codes within a short Hamming distance.
To achieve this aim, we derive the following formula to learn hash codes:
\begin{equation}
\begin{aligned}
\mathop {\min }\limits_{{\bf{Z}} \in {{\left\{ { - 1,1} \right\}}^{r \times n}}} \frac{1}{2}\sum\limits_{{\mathbf{e}}\in {\mathbf{E}}}\sum\limits_{{\hat{\mathbf{x}}}_i,{\hat{\mathbf{x}}}_j \in {\hat{\mathbf{X}}}} \frac{{{h({\hat{\mathbf{x}}}_i,{\mathbf{e}})}{h({\hat{\mathbf{x}}}_j,{\mathbf{e}})}}}{\delta (e)} \\
{{(\frac{{{\mathbf{z}_i}}}{{\sqrt {d({{\hat{\mathbf{x}}}_i})} }} - \frac{{{\mathbf{z}_j}}}{{\sqrt {d({{\hat{\mathbf{x}}}_j})} }})}^2}\\
\end{aligned}
\end{equation}
\begin{equation}
\Rightarrow \mathop {\min }\limits_{\mathbf{Z} \in {\left\{ { -1,1} \right\}^{r \times n}}}\texttt{Tr}(\mathbf{Z}\mathbf{L}_H{\mathbf{Z}^{\texttt{T}}}),
\label{eq21}
\end{equation}
where $\mathbf{L}_H$ denotes the Laplacian matrix of hypergraph.
To reduce the computation complexity,
we avoid explicitly computing the $\mathbf{L}_H$,
and substitute it with the following form
\begin{equation}
\mathbf{L}_H=\mathbf{I}_n - \mathbf{D}_v^{ - 1/2}\mathbf{H}{\mathbf{D}_w}\mathbf{D}_e^{ - 1}{\mathbf{H}^{\texttt{T}}}\mathbf{D}_v^{ - 1/2},
\label{eq22}
\end{equation}
where $\mathbf{D}_v$, $\mathbf{D}_e$ and $\mathbf{D}_w$ are diagonal matrices of vertex degrees,
hyperedge degrees and hyperedge weights, respectively.
Eq.(\ref{eq21}) can be transformed as
\begin{equation}
\mathop {\min }\limits_{\mathbf{Z} \in {\left\{ {-1,1} \right\}^{r \times n}}} - \texttt{Tr}(\mathbf{Z}\mathbf{D}_v^{ - 1/2}\mathbf{H}{\mathbf{D}_w}\mathbf{D}_e^{ - 1}{\mathbf{H}^{\texttt{T}}}\mathbf{D}_v^{ - 1/2}{\mathbf{Z}^{\texttt{T}}}).
\label{eq24}
\end{equation}
As shown in the optimization, this strategy can successfully reduce the computation cost.

\subsubsection{Deep Representation Learning} \label{sub3_2}
Shallow hashing methods with deep features as input still obtain sub-optimal performance
because the separately learned hash codes may not be optimally compatible with the hash function learning.
In this paper, we propose to jointly perform feature learning and hash function learning in a unified deep framework.
We adopt VGG-16 model \cite{VGG16} as our basic deep hash model and
the parameters are initialized with that are pre-trained on the ImageNet dataset  \cite{VGG16}.
We modify the number of neurons in the last fully connected layer as the hash code length
and choose $\texttt{tanh}$ function as the activation function.
$\Phi(\mathbf{x}_i;\Theta)$ is determined as the outputs of the last fully connected layer with data $\mathbf{x}_i$,
where $\Theta$ is the parameters of the deep network model.
$\hat{\Phi}(\mathbf{x}_i;\hat{\Theta})$ represents the feature representation model,
which is the outputs of the second fully connected layer (FC$_{7}$) associated with data $\mathbf{x}_i$,
$\hat{\Theta}$ denotes all the parameters of previous layers before the last fully connected layer.
By encoding images with $\hat{\Phi}(\mathbf{x}_i;\hat{\Theta})$,
we can obtain more powerful deep image representations.
Meanwhile, by minimizing the quantization loss between
the outputs of the deep model $\Phi(\mathbf{x}_i;\Theta)$ and the learned binary codes $\mathbf{z}_i$,
the energy of the learned compact features
can be preserved into the binary features, that is
\begin{equation} \label{eq01}
\mathop {\min }\limits_\Theta \ {\lVert {\Phi(\mathbf{X};\Theta) - \mathbf{Z}} \rVert _\texttt{F}^2}.
\end{equation}
\subsubsection{Overall Objective Function}
By comprehensively considering the above factors,
we derive the following objective formulation
\begin{equation}
\begin{aligned}
\mathop {\min }\limits_{\mathbf{Z} \in {\left\{ { -1,1} \right\}^{r \times n}},\mathbf{P}, \Theta} \  & \rVert {\mathbf{Z} - {\mathbf{P}^{\texttt{T}}}\mathbf{Y}} \rVert_{2,1}
-\alpha \texttt{Tr}(\mathbf{Z}\mathbf{V}\mathbf{\Lambda}^{-1} {\mathbf{V}^{\texttt{T}}}{\mathbf{Z}^{\texttt{T}}}) \\
-& \beta \texttt{Tr}(\mathbf{Z}\mathbf{D}_v^{ - 1/2}\mathbf{H}{\mathbf{D}_w}\mathbf{D}_e^{ - 1}{\mathbf{H}^{\texttt{T}}}\mathbf{D}_v^{ - 1/2}{\mathbf{Z}^{\texttt{T}}}) \\
+& \nu  \ {\lVert {\Phi(\mathbf{X};\Theta) - \mathbf{Z}}  \rVert _\texttt{F}^2}.
\end{aligned}
\label{eq34}
\end{equation}
Here $\alpha$, $\beta$ and $\nu$ are used to balance the effect of different terms.
By joint learning, the visual similarity preservation and dual-level semantic transfer
can steer the feature representation and hash function learning.
On the other hand,
the updated deep representation and hash model
can further benefit the similarity preservation and semantic transfer.

\subsection{Efficient Discrete Optimization}
Optimizing the hash codes in Eq.(\ref{eq34}) is actually NP-hard problem
due to the discrete constraint and $\ell_{2,1}$-norm imposed on the hash codes.
Existing discrete optimization methods cannot be directly applied for solving our problem.
In this paper, we propose an efficient discrete optimization solution based on
Augmented Lagrangian Multiplier (ALM) \cite{ALM},
which keeps the binary constraints $\mathbf{Z} \in {\{-1,1\}} ^{r \times n}$ and
solves the hash codes directly.
Our idea is introducing auxiliary variables to move the discrete constraints out of $\ell_{2,1}$-norm,
and transform the optimization problem to an equivalent one that can be tackled more easily.
First, auxiliary variables $\mathbf{A}$ and $\mathbf{B}$ are added as
$\mathbf{A} = \mathbf{Z} - {\mathbf{P}^{\texttt{T}}\mathbf{Y}}, \ \mathbf{B}=\mathbf{Z}$.
With the substitution of variables,
the objective function is then reformulated as
\begin{equation}
\begin{aligned}
& \mathop {\min }\limits_{ \mathbf{Z} \in {\left\{ { -1,1} \right\}^{r \times n}},\mathbf{P},\mathbf{A},\mathbf{B}, \Theta}
 {\lVert \mathbf{A} \rVert_{2,1}} + \frac{\mu }{2}\rVert {\mathbf{Z} - {\mathbf{P}^{\texttt{T}}}\mathbf{Y} - \mathbf{A} +
\frac{\mathbf{E}_A}{\mu }} \rVert_\texttt{F}^2 \\
& + \frac{\mu }{2}\lVert {\mathbf{Z} - \mathbf{B} + \frac{\mathbf{E}_B}{\mu }} \rVert_\texttt{F}^2
 -\alpha \texttt{Tr}(\mathbf{B}\mathbf{V}\mathbf{\Lambda}^{-1} {\mathbf{V}^{\texttt{T}}}{\mathbf{Z}^{\texttt{T}}}) \\
& -\beta \texttt{Tr}(\mathbf{B}\mathbf{D}_v^{ - 1/2}\mathbf{H}{\mathbf{D}_w}\mathbf{D}_e^{ - 1}{\mathbf{H}^{\texttt{T}}}\mathbf{D}_v^{ - 1/2}{\mathbf{Z}^{\texttt{T}}}) \\
&+\nu  \ {\lVert {\Phi(\mathbf{X};\Theta) - \mathbf{Z}} \rVert _\texttt{F}^2},
\end{aligned}
\label{eq35}
\end{equation}
where ${\mathbf{E}_A}$ and ${\mathbf{E}_B}$
measure the difference between the target and auxiliary variables,
$\mu$ adjusts the balance between the regularization terms.
With the transformation,
we can drive the following iterative optimization steps to solve Eq.(\ref{eq35}).

\textbf{-Update} $\mathbf{A}$. The optimization formula for $\mathbf{A}$ is
\begin{equation}
\mathop {\min }\limits_{\mathbf{A}}{\lVert \mathbf{A} \rVert_{2,1}} + \frac{\mu }{2}\lVert {\mathbf{Z} - {\mathbf{P}^{\texttt{T}}}\mathbf{Y} - \mathbf{A} +
\frac{\mathbf{E}_A}{\mu }} \rVert_\texttt{F}^2.
\label{a1}
\end{equation}
The Eq.(\ref{a1}) can be written in the following form
\begin{equation}
\mathop {\min }\limits_{\mathbf{A}} \frac{1}{2}\lVert {\mathbf{A} - \mathbf{T}} \rVert_\texttt{F}^2
+ \frac{1}{\mu }{\lVert \mathbf{A} \rVert_{2,1}},
\label{a2}
\end{equation}
where
$\mathbf{T} =  \mathbf{Z} - {\mathbf{P}^{\texttt{T}}}\mathbf{Y} +  \frac{\mathbf{E}_A}{\mu }$.
The solution of $\mathbf{A}$ is
\begin{eqnarray}
\ \mathbf{A} \left( {:,i} \right) = \left\{ {\begin{array}{*{20}{l}}
{\frac{\lVert {{t_i}} \rVert - \frac{1}{\mu }}{{\lVert {{t_i}} \rVert}}{t_i},}&{if \ \ \frac{1}{\mu } < \lVert {{t_i}} \rVert}\\
{0,}&{otherwise.}
\end{array}} \right.\
\label{a4}
\end{eqnarray}

\textbf{-Update} $\mathbf{P}$.
Similarly, the optimization formula for $\mathbf{P}$ is
\begin{equation}\label{p1}
\mathop {\min }\limits_\mathbf{P}
\frac{\mu }{2}\lVert {\mathbf{Z} - {\mathbf{P}^{\texttt{T}}}\mathbf{Y} - \mathbf{A} + \frac{\mathbf{E}_A}{\mu }} \rVert_\texttt{F}^2.
\end{equation}
By calculating the derivative of Eq.(\ref{p1}) w.r.t $\mathbf{P}$ and setting it to 0, we can obtain that
\begin{equation} \label{p2}
\mathbf{P} = {(\mathbf{Y}{\mathbf{Y}^{\texttt{T}}})^{ - 1}}\mathbf{Y}({\mathbf{Z}^{\texttt{T}} - \mathbf{A}^{\texttt{T}} + \frac{\mathbf{E}_A}{\mu } }).
\end{equation}

\textbf{-Update} $\mathbf{B}$.
The optimization formula for $\mathbf{B}$ becomes
\begin{equation}\label{b1}
\begin{aligned}
\mathop {\min }\limits_{\mathbf{B}}
& \frac{\mu }{2}\lVert {\mathbf{Z} - \mathbf{B} + \frac{\mathbf{E}_B}{\mu }} \rVert_\texttt{F}^2
-\alpha \texttt{Tr}(\mathbf{B}\mathbf{V}\mathbf{\Lambda}^{-1} {\mathbf{V}^{\texttt{T}}}{\mathbf{Z}^{\texttt{T}}}) \\
& -\beta \texttt{Tr}(\mathbf{B}\mathbf{D}_v^{ - 1/2}\mathbf{H}{\mathbf{D}_w}\mathbf{D}_e^{ - 1}{\mathbf{H}^{\texttt{T}}}\mathbf{D}_v^{ - 1/2}{\mathbf{Z}^{\texttt{T}}}). \\
\end{aligned}
\end{equation}
By calculating the derivative of Eq.(\ref{b1}) w.r.t $\mathbf{B}$ and setting it to 0, we derive that
\begin{equation}\label{b2}
\mathbf{B} = \mathbf{Z} + \frac{\mathbf{E}_B}{\mu }
+ \frac{\alpha}{\mu } {\mathbf{Z}\mathbf{V}\mathbf{\Lambda}^{-1} {\mathbf{V}^{\texttt{T}}}}
+ \frac{\beta}{\mu } {\mathbf{Z}\mathbf{D}_v^{ - 1/2}\mathbf{H}{\mathbf{D}_w}\mathbf{D}_e^{-1}{\mathbf{H}^{\texttt{T}}}\mathbf{D}_v^{ - 1/2}}.
\end{equation}

\textbf{-Update} $\mathbf{Z}$ and $\Theta$.
The optimization formula for $\mathbf{Z}$ is
\begin{equation}\label{z1}
\begin{aligned}
&\mathop {\min }\limits_{\mathbf{Z} \in {\left\{ { -1,1} \right\}^{r \times n}}}
\ \frac{\mu }{2}\rVert {\mathbf{Z} - {\mathbf{P}^{\texttt{T}}}\mathbf{Y} - \mathbf{A} +
\frac{\mathbf{E}_A}{\mu }} \rVert_\texttt{F}^2 \\
& + \ \frac{\mu }{2}\lVert {\mathbf{Z} - \mathbf{B} + \frac{\mathbf{E}_B}{\mu }} \rVert_\texttt{F}^2
-\alpha \texttt{Tr}(\mathbf{B}\mathbf{V}\mathbf{\Lambda}^{-1} {\mathbf{V}^{\texttt{T}}}{\mathbf{Z}^{\texttt{T}}}) \\
& -\beta \texttt{Tr}(\mathbf{B}\mathbf{D}_v^{ - 1/2}\mathbf{H}{\mathbf{D}_w}\mathbf{D}_e^{ - 1}{\mathbf{H}^{\texttt{T}}}\mathbf{D}_v^{ - 1/2}{\mathbf{Z}^{\texttt{T}}}) \\
& +\nu  \ {\lVert {\Phi(\mathbf{X};\Theta) - \mathbf{Z}} \rVert _\texttt{F}^2}.
\end{aligned}
\end{equation}
Eq.(\ref{z1}) can be transformed to
\begin{equation}\label{z2}
\begin{aligned}
& \mathop {\min }\limits_{\mathbf{Z} \in {\left\{ { -1,1} \right\}^{r \times n}}}
-\texttt{Tr}({\mathbf{Z}^{\texttt{T}}}(\mu \mathbf{A} + \mu {\mathbf{P}^{\texttt{T}}}\mathbf{Y}+ \mu \mathbf{B}
- {\mathbf{E}_A} - {\mathbf{E}_B} \\
& +\alpha \mathbf{B}\mathbf{V}\mathbf{\Lambda}^{-1} {\mathbf{V}^{\texttt{T}}}
  +\beta \mathbf{B}\mathbf{D}_v^{ - 1/2}\mathbf{H}{\mathbf{D}_w}\mathbf{D}_e^{ - 1}{\mathbf{H}^{\texttt{T}}}\mathbf{D}_v^{ - 1/2} \\
& + 2 \nu \Phi(\mathbf{X};\Theta))).
\end{aligned}
\end{equation}

The solution of $\mathbf{Z}$ can be obtained by the following closed form
\begin{equation}\label{z3}
\begin{aligned}
{\mathbf{Z}} =
\texttt{sgn}(&\mu \mathbf{A} + \mu {\mathbf{P}^{\texttt{T}}}\mathbf{Y}+ \mu \mathbf{B}
- {\mathbf{E}_A} - {\mathbf{E}_B}
+\alpha \mathbf{B}\mathbf{V}\mathbf{\Lambda}^{-1} {\mathbf{V}^{\texttt{T}}} \\
+ &\beta \mathbf{B}\mathbf{D}_v^{ - 1/2}\mathbf{H}{\mathbf{D}_w}\mathbf{D}_e^{ - 1}{\mathbf{H}^{\texttt{T}}}\mathbf{D}_v^{ - 1/2}
+ 2 \nu \Phi(\mathbf{X};\Theta)).
\end{aligned}
\end{equation}
As shown in Eq.(\ref{z3}),
the hash codes $\mathbf{Z}$ are directly solved with a single hash code-solving operation,
which is faster than iterative hash codes solving step in DCC.

After that, the learned $\mathbf{Z}$ is taken into the
Subsection \emph{\ref{sub3_2}. Deep Representation Learning}
to update the neural network parameters $\Theta$ through the standard backpropagation
and stochastic gradient descent.

\textbf{-Update} ${\mathbf{E}_A}$, ${\mathbf{E}_B}$ and $\mu$.
The ALM parameters are updated by
\begin{equation}\label{alm1}
\begin{aligned}
{\mathbf{E}_A}  &= {\mathbf{E}_A}  + \mu ({\mathbf{Z} - {\mathbf{P}^{\texttt{T}}}\mathbf{Y} - \mathbf{A}} ),\\
{\mathbf{E}_B} &= {\mathbf{E}_B} + \mu (\mathbf{Z} - \mathbf{B}),\\
\mu  &= \rho \mu.
\end{aligned}
\end{equation}
Here $\rho$ controls the convergence speed.

{
The key steps of proposed DSTDH are illustrated in Algorithm \ref{alg1}.}
\begin{algorithm}[t]
\caption{\textbf{Key steps of DSTDH}}
\label{alg1}
\begin{algorithmic}[1]
\Require
Training image matrix $\mathbf{X} \in\mathbb{R}^{d\times n} $,
image-tag relation matrix $\mathbf{Y} \in\mathbb{R}^{c\times n} $,
the number of anchor points $m$,
the number of concepts $a$,
hash code length $r$,
parameter $\alpha,\beta$ and $\nu$.
\Ensure
Deep hash function $h(\bm{{\rm x}})$.
    \State Initialize the deep model parameters $\Theta$ by the pre-trained VGG-16 model on ImageNet.
    \State Randomly initialize $\mathbf{Z}=\mathbf{B}$ as $\{-1, 1\}^{r\times n}$.
    \State Initialize $\mathbf{A}$, $\mathbf{P}$, $\mathbf{E}_A$ and $\mathbf{E}_B$ as the matrices with all elements as 0.
\Repeat
        \State Update $\mathbf{A}$ by Eq.(\ref{a4}).
        \State Update $\mathbf{P}$ by Eq.(\ref{p2}).
        \State Update $\mathbf{B}$ by Eq.(\ref{b2}).
        \State Update $\mathbf{Z}$ and $\Theta$ by Eq.(\ref{z3}) and Eq.(\ref{eq01}).
        \State Update $\mathbf{E}_A, \mathbf{E}_B$ and $\mu$ by Eq.(\ref{alm1}).
\Until{convergence.}
\State \textbf{Return} $h(\bm{{\rm x}})= \texttt{sgn}(\Phi(\mathbf{X};\Theta))$.
\end{algorithmic}
\end{algorithm}

\subsection{Discussions}
\subsubsection{Convergency Analysis}
In the discrete optimization process,
when other variables are fixed, the objective function Eq.(\ref{eq35}) is convex for one variable.
Therefore, the value of the objective function
will remain unchanged or monotonically decrease in each iterative optimization processes,
and finally reaches the local minimum after several iterations.
In addition, the experimental results in Section \ref{exper} also validate the convergence of our approach.

\subsubsection{Computation Complexity Analysis}
\begin{table}
\caption{Statistics of the experimental datasets}
\centering
\setlength{\tabcolsep}{5mm}{
\begin{tabular}{lcc}
  \hline
  Datasets &MIR Flickr &NUS-WIDE\\
  \hline
  \#Categories & 24      & 21     \\
  \#Retrieval Set  & 17,749 & 193,749 \\
  \#Query Set      & 2,266  & 2,085  \\
  \#Training Set   & 5,325  & 5,000 \\
  \hline
  Tag Feature (BOW)      &1,386-D &1,000-D  \\
  Hand-crafted Image Feature &1,000-D &500-D  \\
  (BOVW) \cite{BOVW} & &  \\
  Deep Image Feature (VGG)       &4,096-D & 4,096-D \\
  \hline
\end{tabular}
\label{table2}}
\end{table}
The computation complexity of visual anchor graph construction and hypergraph construction
are $O(dmn)$ + $O(dan)$.
At each iteration of  discrete optimization,
it takes $O(rcn)$ for updating $\mathbf{A}$,
$O({c^3} + rcn + 2{c^2}n)$ for updating $\mathbf{P}$,
$O((m + a)rn)$ for updating $\mathbf{B}$,
and $O((m + a)rn)$ for updating $\mathbf{Z}$.
Finally, the computation complexity of the discrete optimization process is $O(tn)$,
where \emph{t} is the number of iterations and $t\ll n$.
Since the computation complexity of our method is linearly related to the number of images in the social image dataset,
the fast social image retrieval can be achieved.

\section{Experiments}\label{exper}
\subsection{Datasets and Evaluation Metrics}
\begin{table}
\centering
\caption{Comparison results with unsupervised hashing on MIR Flickr.
The best result in each column is marked with bold. The below is the same.}
\setlength{\tabcolsep}{2mm}{
\begin{tabular}{c|c|c|c|c|c}
\hline
{Methods} & {Feature}& \multicolumn{4}{c}{MAP}  \\
\cline{3-6} & & 16 bits & 32 bits & 64 bits & 128 bits  \\ \hline \hline
LSH      & BOVW &	0.5830	&	0.5830	&	0.5827	&	0.5839	\\
SKLSH    & BOVW &	0.5791	&	0.5896	&	0.5895	&	0.5939	\\
PCAH     & BOVW &	0.5822	&	0.5825	&	0.5821	&	0.5832	\\
SH       & BOVW &	0.5806	&	0.5854	&	0.5878	&	0.5903	\\
SGH      & BOVW &	0.5871	&	0.5847	&	0.5844	&	0.5837	\\
ITQ      & BOVW &	0.5820	&	0.5826	&	0.5820	&	0.5824	\\
LSMH     & BOVW &	0.5821	&	0.5827	&	0.5815	&	0.5828	\\
SADH-L   & BOVW &	0.5823	&	0.5822	&	0.5825	&	0.5809	\\
DSTDH-L & BOVW &	\textbf{0.6111}	&	\textbf{0.6291}	&	\textbf{0.6355}	&	\textbf{0.6389}	\\ \hline
\hline
LSH         & VGG &	0.6103	&	0.6372	&	0.6513	&	0.6736	\\
SKLSH       & VGG &	0.5823	&	0.5824	&	0.5826	&	0.5828	\\
PCAH        & VGG &	0.6603	&	0.6519	&	0.6418	&	0.6320	\\
SH          & VGG &	0.6510	&	0.6373	&	0.6292	&	0.6288	\\
SGH         & VGG &	0.5841	&	0.5850	&	0.5893	&	0.5958	\\
ITQ         & VGG &	0.7308  &   0.7392  &   0.7413  &   0.7445  \\
LSMH        & VGG &	0.7079	&	0.7054	&	0.6979	&	0.6739	\\
SADH-L      & VGG &	0.6182	&	0.6058	&	0.6019	&	0.5985	\\
DSTDH-L    & VGG &	\textbf{0.7388}	&	\textbf{0.7436}	&	\textbf{0.7481}	&	\textbf{0.7515}\\ \hline
\hline
UH-BDNN     & VGG &	0.7215	&	0.7028	&	0.6854	&	0.6715	\\
DeepBit     & Raw &	0.6134	&	0.6328	&	0.6527	&	0.6722	\\
SADH        & Raw &	0.7477	&	0.7458	&	0.7565	&	0.7601	\\
GreedyHash  & Raw &	0.6780	&	0.6889	&	0.7131	&	0.7287	\\
DSTDH     & Raw &	\textbf{0.7664} &	\textbf{0.7769} &	\textbf{0.7893}	&	\textbf{0.7980}\\\hline
\end{tabular}}\label{table4}
\end{table}
\begin{table}
\centering
\caption{{Comparison results with unsupervised hashing on NUS-WIDE.}}
\setlength{\tabcolsep}{2mm}{
\begin{tabular}{c|c|c|c|c|c}
\hline
{Methods} & {Feature}& \multicolumn{4}{c}{MAP}  \\
\cline{3-6} & & 16 bits & 32 bits & 64 bits & 128 bits  \\ \hline \hline
LSH      & BOVW &	0.3634 	&	0.3642 	&	0.3650 	&	0.3648 	\\
SKLSH    & BOVW &	0.3555 	&	0.3560 	&	0.3572 	&	0.3567 	\\
PCAH     & BOVW &	0.3642 	&	0.3645 	&	0.3647 	&	0.3642 	\\
SH       & BOVW &	0.3561 	&	0.3573 	&	0.3526 	&	0.3524 	\\
SGH      & BOVW &	0.3668 	&	0.3662 	&	0.3664 	&	0.3657 	\\
ITQ      & BOVW &	0.3640 	&	0.3641 	&	0.3643 	&	0.3637 	\\
LSMH     & BOVW &	0.3640 	&	0.3640 	&	0.3646 	&	0.3635 	\\
SADH-L   & BOVW &	0.3638 	&	0.3639 	&	0.3643 	&	0.3649 	\\
DSTDH-L & BOVW &	\textbf{0.4126}	&	\textbf{0.4272}	&	\textbf{0.4464}	&	\textbf{0.4604}	\\ \hline
\hline																
LSH      & VGG &	0.4242	&	0.4572	&	0.4862	&	0.5085	\\
SKLSH    & VGG &	0.3632	&	0.3637	&	0.3641	&	0.3644	\\
PCAH     & VGG &	0.5234	&	0.5053	&	0.4889	&	0.4821	\\
SH       & VGG &	0.4401	&	0.4674	&	0.4578	&	0.4980	\\
SGH      & VGG &	0.3652	&	0.3679	&	0.3722	&	0.3823	\\
ITQ      & VGG &	0.5808	&	0.5965	&	0.5998	&	0.6116	\\
LSMH     & VGG &	0.5607	&	0.5639	&	0.5526	&	0.5218	\\
SADH-L   & VGG &	0.4201	&	0.4152	&	0.4043	&	0.3988	\\
DSTDH-L & VGG &	\textbf{0.5843}	&	\textbf{0.6032}	&	\textbf{0.6099}	&	\textbf{0.6150}	\\ \hline
\hline																
UH-BDNN    & VGG &	0.5609	&	0.5445	&	0.5221	&	0.5180	\\
DeepBit    & Raw &	0.4209	&	0.4264	&	0.4587	&	0.5224	\\
SADH       & Raw &	0.5722	&	0.5919	&	0.6049	&	0.6002	\\
GreedyHash & Raw &	0.4921	&	0.5463	&	0.5765	&	0.5914	\\
DSTDH     & Raw &\textbf{0.6324}&\textbf{0.6442}&\textbf{0.6526}&\textbf{0.6551}	\\\hline
\end{tabular}}\label{table5}
\end{table}
We adopt two publicly available social image retrieval datasets to conduct experiments,
MIR Flickr \cite{MIRFlickr} and NUS-WIDE \cite{NUS}.
They are both downloaded from Flickr.
Table \ref{table2} summarizes the basic statistics of these two datasets.

We evaluate the social image retrieval performance by
Mean Average Precision (MAP) \cite{MAP1,MAP2} and precision-recall curve \cite{LSMH}.
The labels are adopted as the groundtruth for evaluation.
According to whether two images share at least one identical label, they can be judged as semantically similar or not.
For the two evaluation protocols, the higher value indicates better performance.
\begin{figure*}[]
\begin{minipage}[t]{0.24\textwidth}
\makeatletter
\def\fnum@figure{}
\makeatother
\includegraphics[scale=.4]{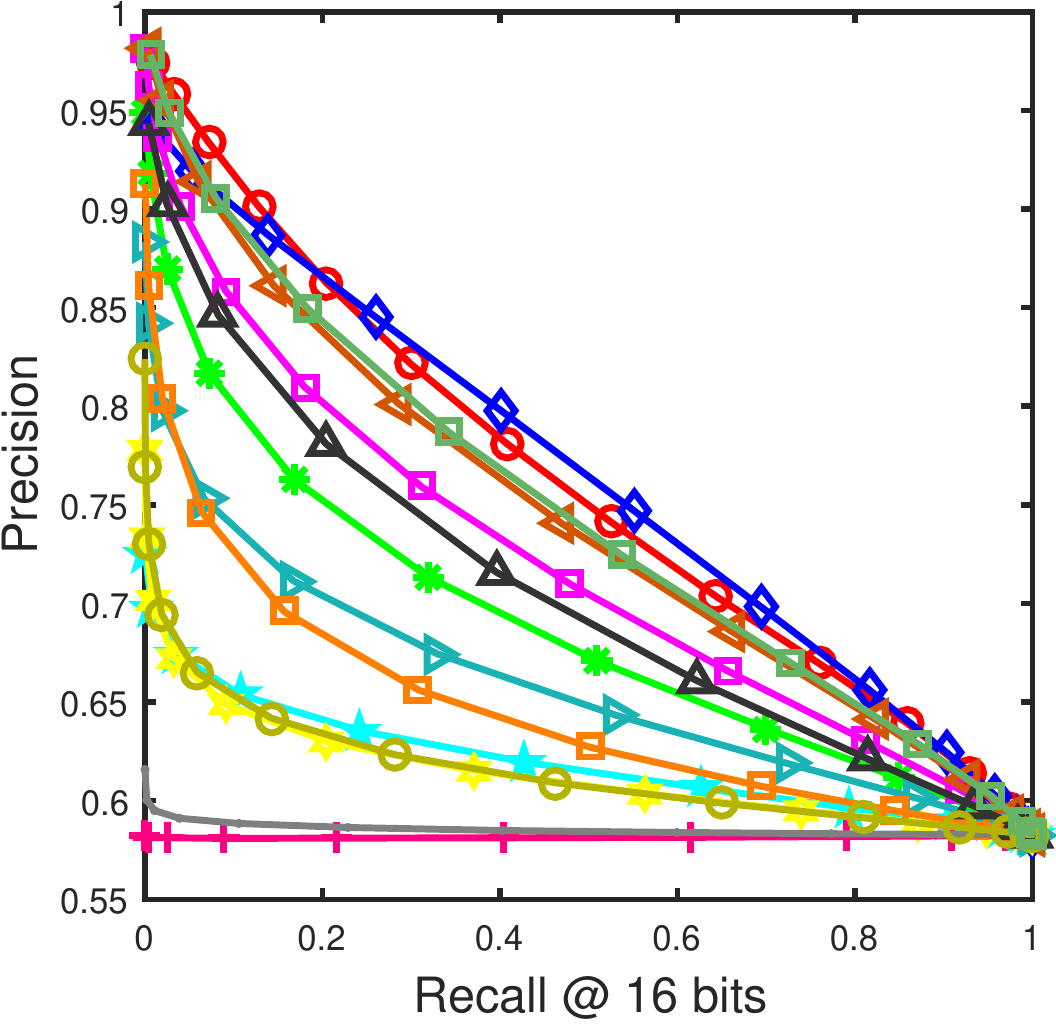}
\end{minipage}
\begin{minipage}[t]{0.24\textwidth}
\makeatletter
\def\fnum@figure{}
\makeatother
\includegraphics[scale=.4]{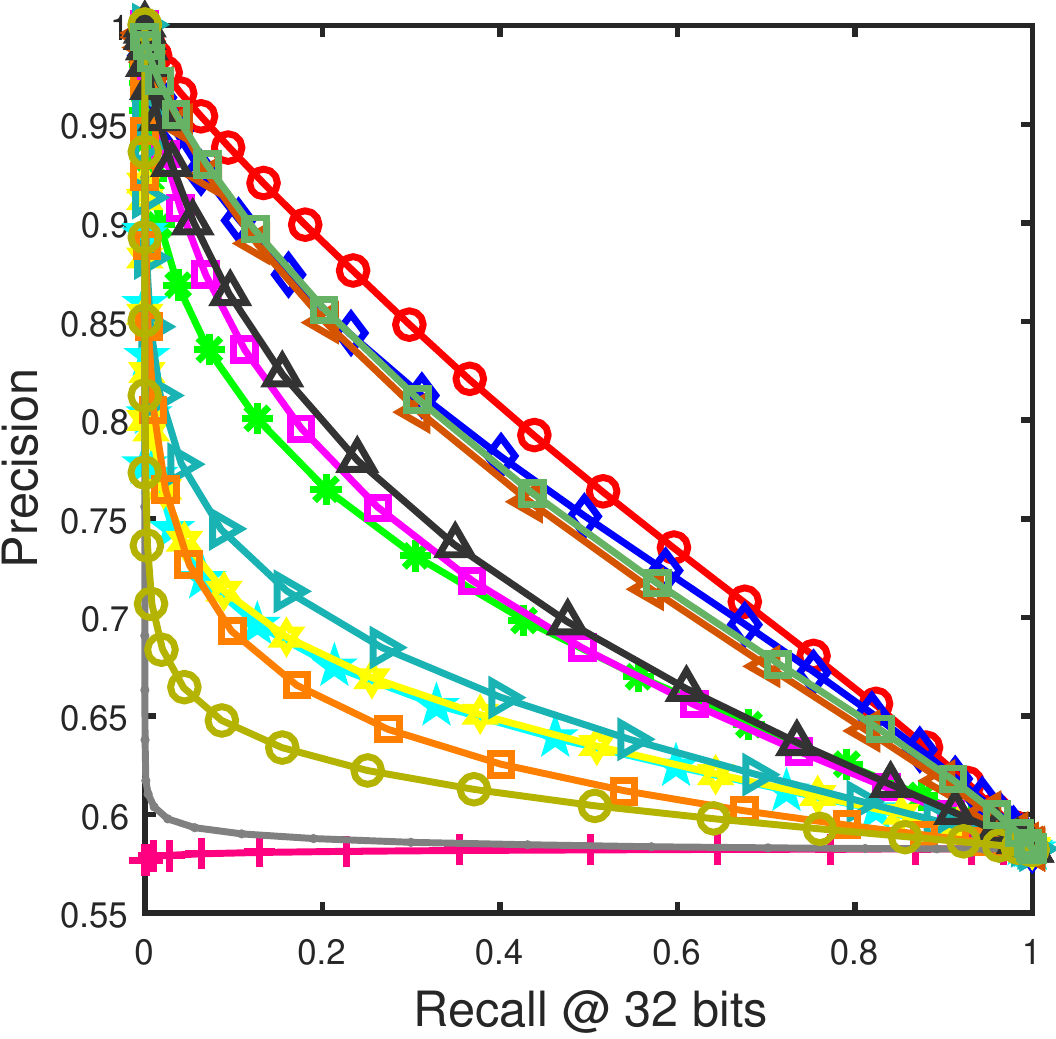}
\end{minipage}
\begin{minipage}[t]{0.24\textwidth}
\makeatletter
\def\fnum@figure{}
\makeatother
\includegraphics[scale=.4]{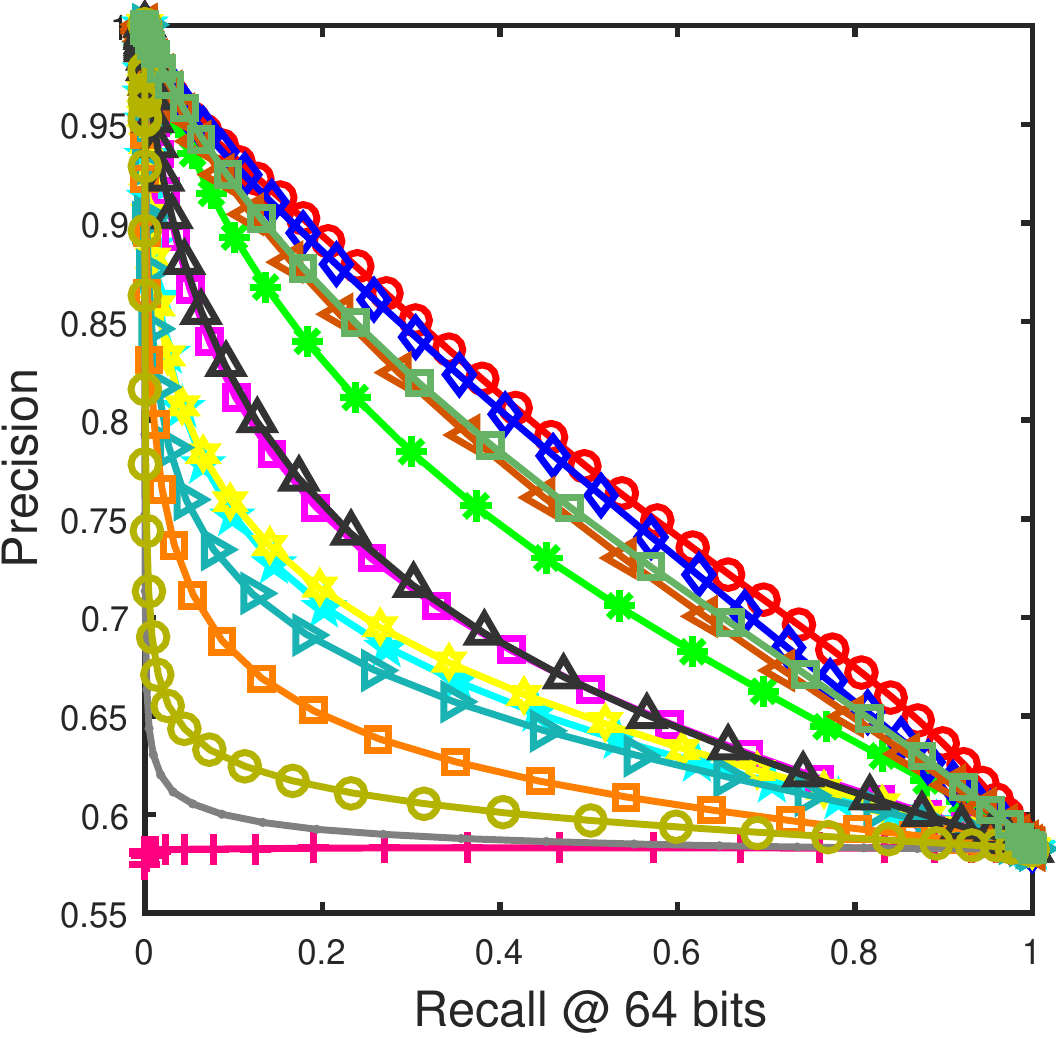}
\end{minipage}
\begin{minipage}[t]{0.24\textwidth}
\centering
\makeatletter
\def\fnum@figure{}
\makeatother
\includegraphics[scale=.4]{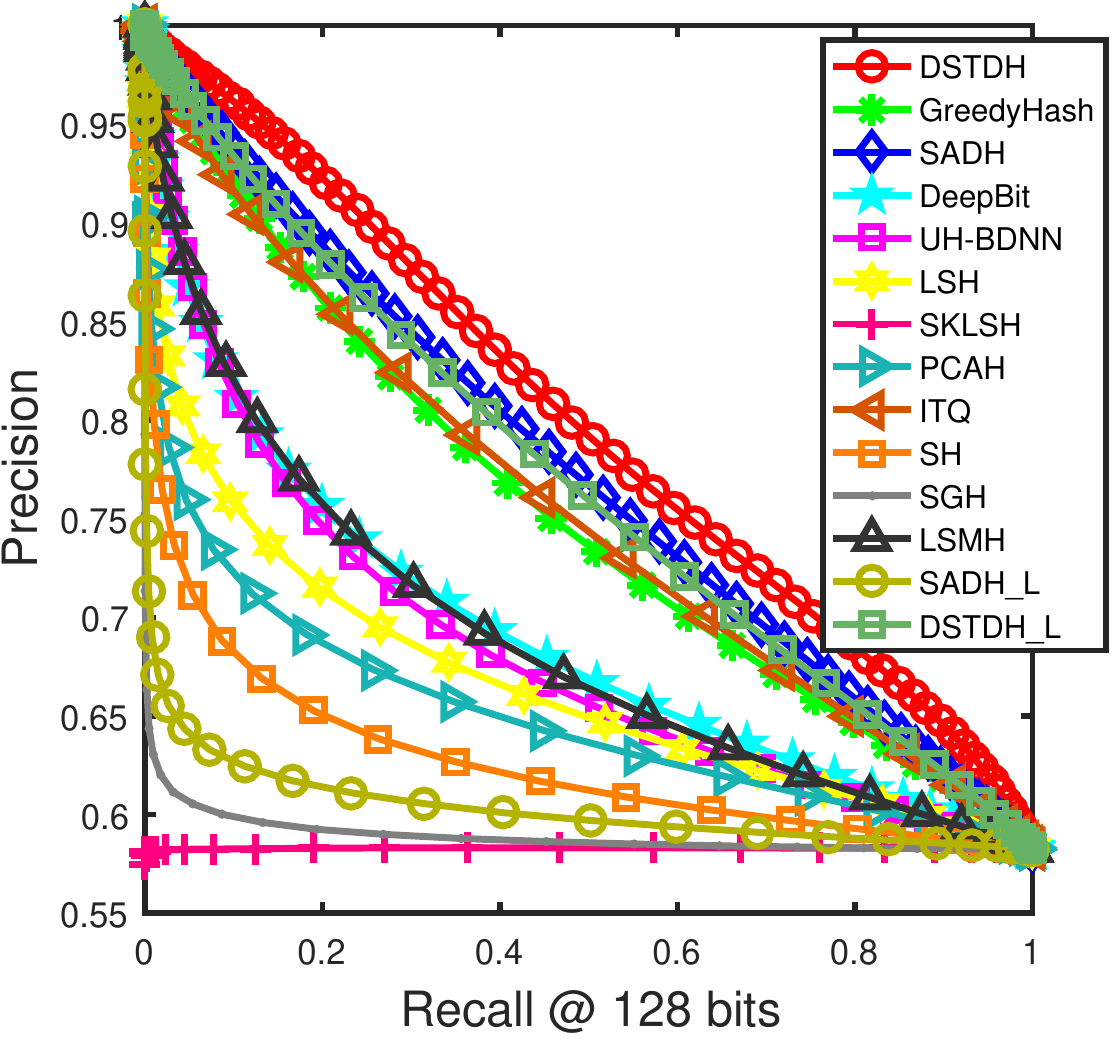}
\end{minipage}
\caption{{The precision-recall curves on MIR Flickr.}}
\label{FigPR_mir}
\end{figure*}
\begin{figure*}[]
\begin{minipage}[t]{0.24\textwidth}
\makeatletter
\def\fnum@figure{}
\makeatother
\includegraphics[scale=.4]{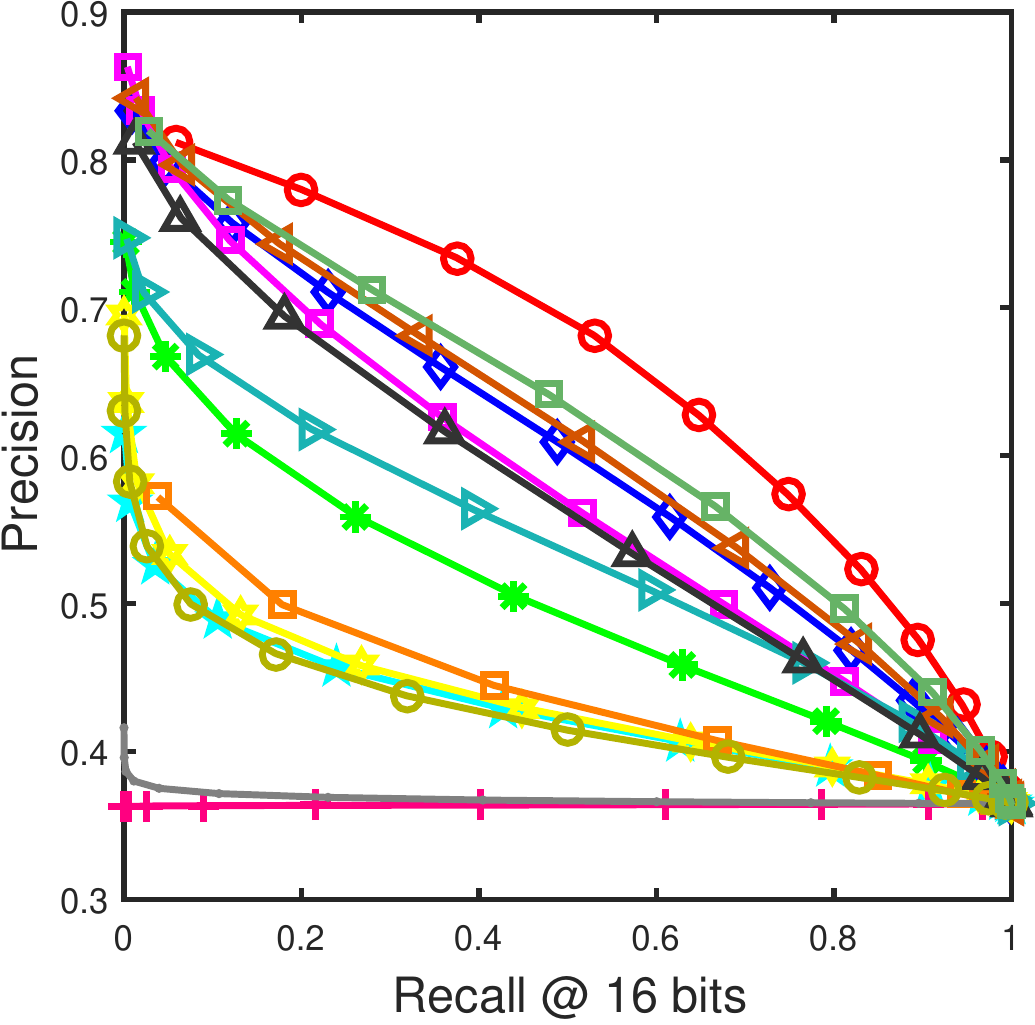}
\end{minipage}
\begin{minipage}[t]{0.24\textwidth}
\makeatletter
\def\fnum@figure{}
\makeatother
\includegraphics[scale=.4]{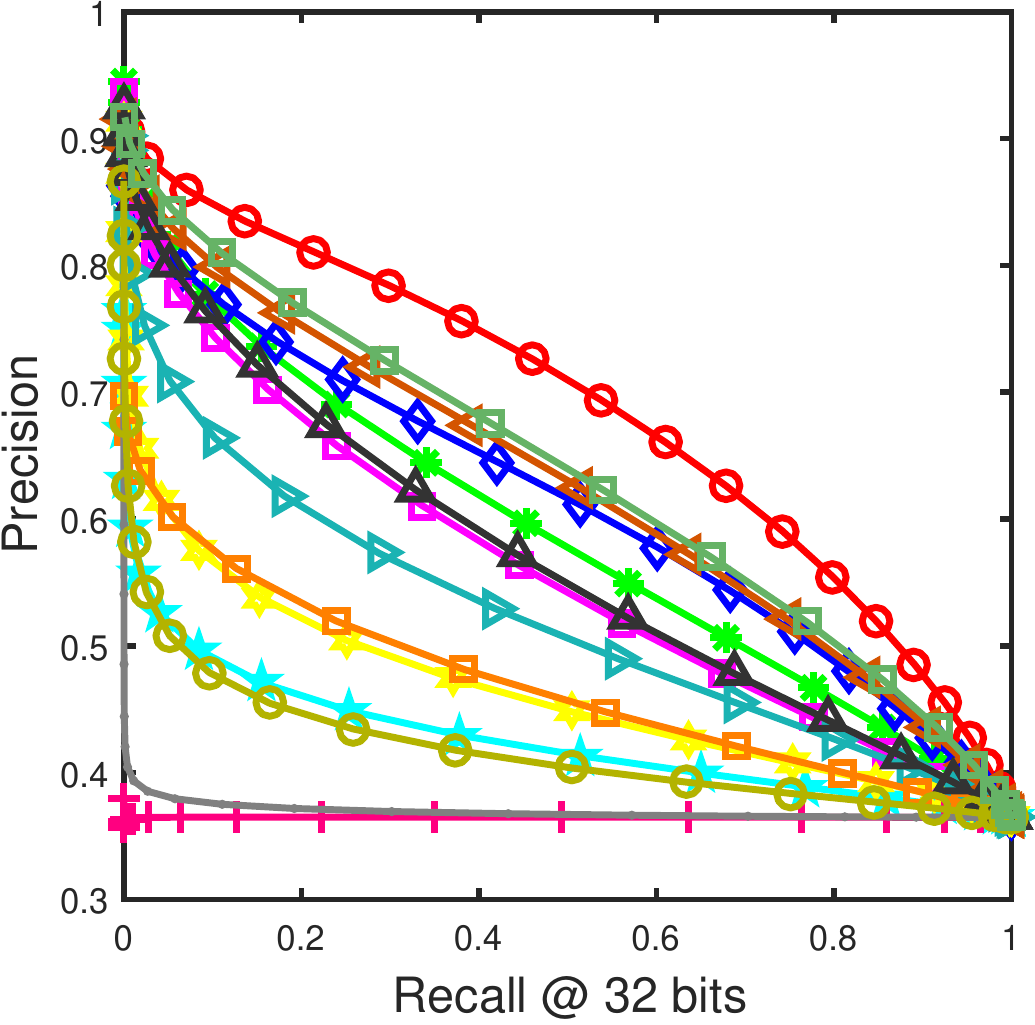}
\end{minipage}
\begin{minipage}[t]{0.24\textwidth}
\makeatletter
\def\fnum@figure{}
\makeatother
\includegraphics[scale=.4]{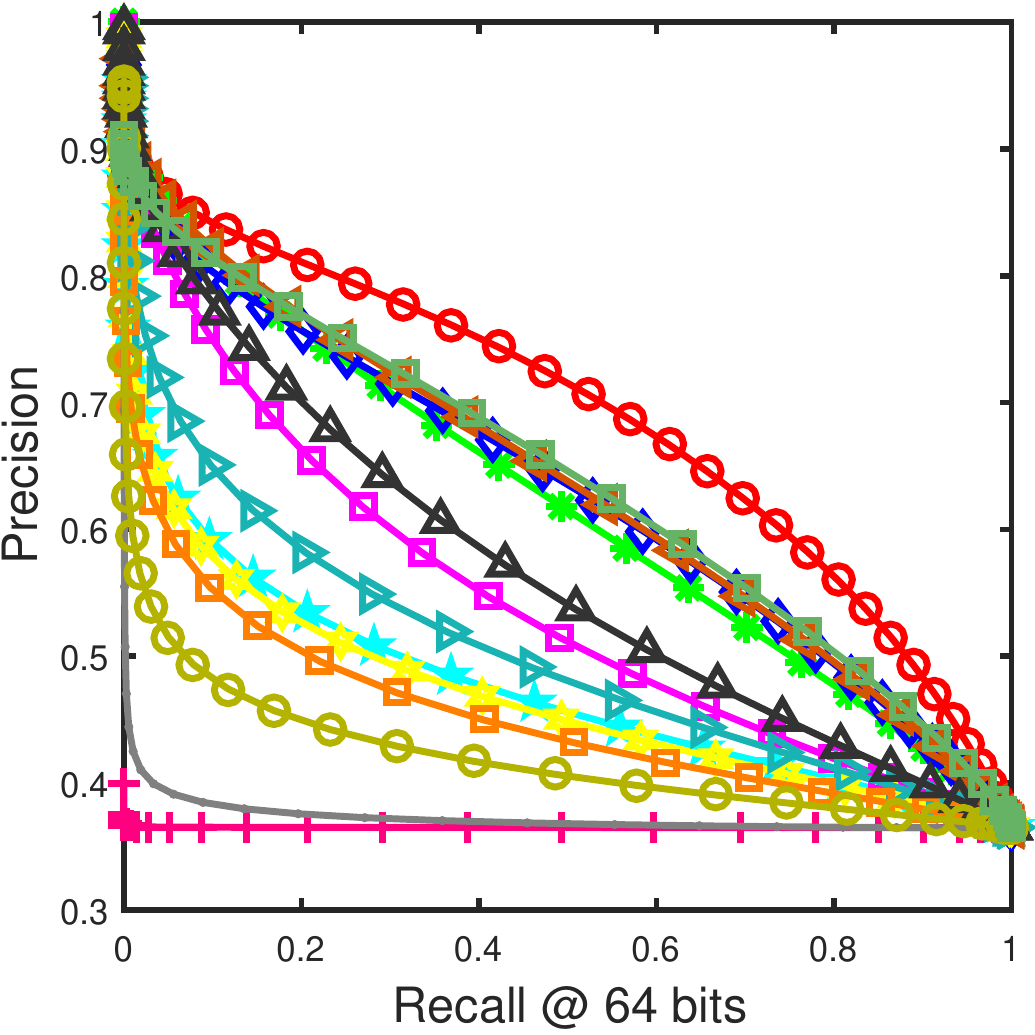}
\end{minipage}
\begin{minipage}[t]{0.24\textwidth}
\centering
\makeatletter
\def\fnum@figure{}
\makeatother
\includegraphics[scale=.4]{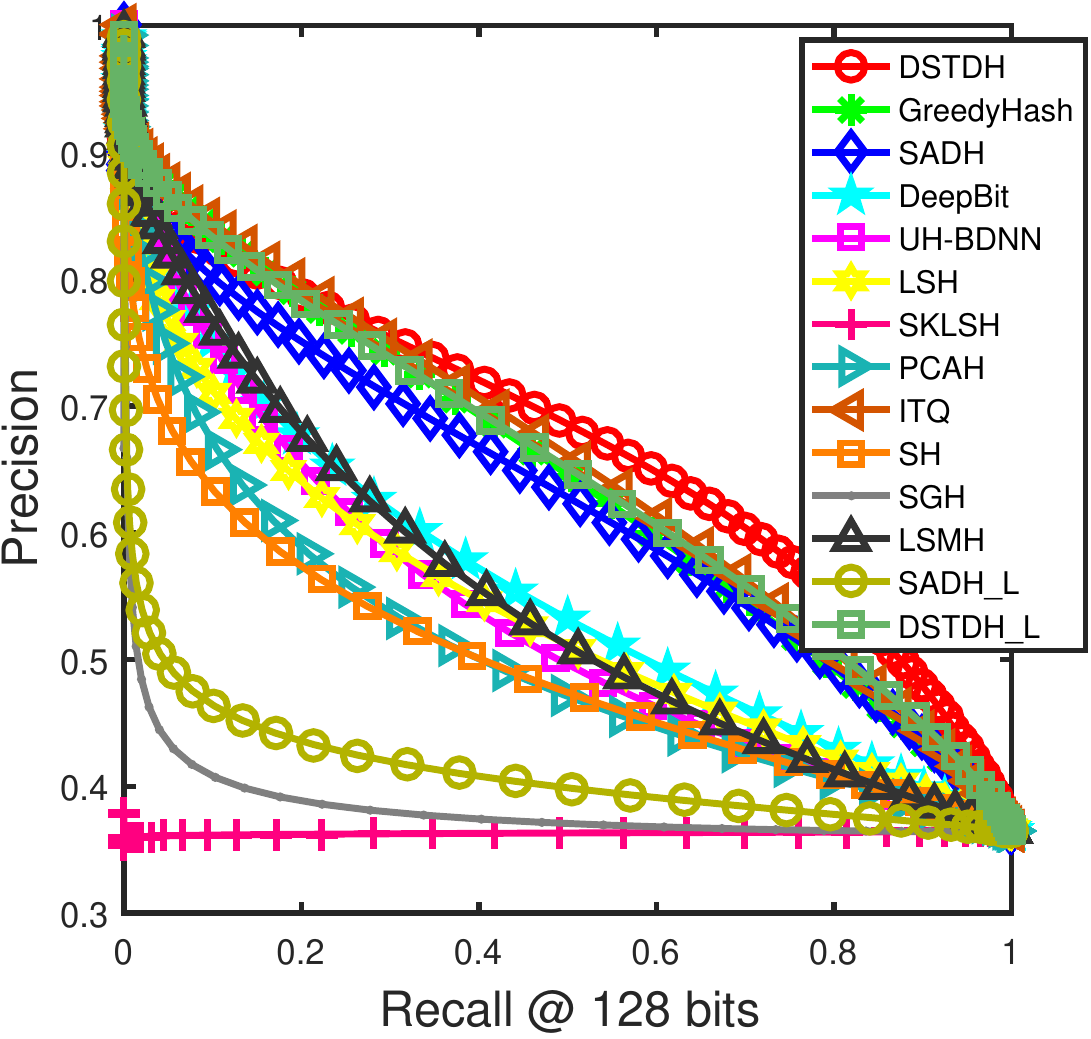}
\end{minipage}
\caption{{The precision-recall curves on NUS-WIDE.}}
\label{FigPR_nus}
\end{figure*}
\subsection{Evaluation Baselines}
We compare DSTDH method and its linear variant DSTDH-L with several state-of-the-art unsupervised hashing methods.
DSTDH-L employs the linear hash function $h(\mathbf{x}) = \mathbf{z} = \texttt{sgn}(\mathbf{W}_l^{\texttt{T}}\mathbf{x})$.
It learns the hash functions with linear regression model,
which minimizes the differences between the generated binary codes and the projected continuous ones by
$\mathop {\min }\limits_{{\mathbf{W}_l}} \left\| {\mathbf{Z} - {\mathbf{W}_l^{\texttt{T}}}\mathbf{X}} \right\|_\texttt{F}^2$.
Then we can get
$\mathbf{W}_l = (\mathbf{XX}^{\texttt{T}})^{-1}\mathbf{XZ}^{\texttt{T}}$.
The baseline methods are divided into four categories:
\begin{itemize}
  \item Eight unsupervised shallow hashing methods using the handcrafted feature:
  Locality Sensitive Hashing (LSH) \cite{LSH},
  Locality-Sensitive Hashing with Shift-invariant Kernels (SKLSH) \cite{SKLSH},
  PCA Hashing (PCAH) \cite{PCAH},
  Spectral Hashing (SH) \cite{SH},
  Scalable Graph Hashing (SGH) \cite{SGH},
  Iterative Quantization (ITQ) \cite{ITQ},
  Latent Semantic Minimal Hashing (LSMH) \cite{LSMH} and
  Similarity-Adaptive Deep Hashing with linear hash function (SADH-L) \cite{SADH}.
  \item Aforementioned unsupervised shallow hashing methods using 4,096-dimensional deep feature extracted by our feature learning model (we denote it as VGG feature for clarity).
  \item Four unsupervised deep hashing methods:
  Unsupervised Deep Learning of Compact Binary Descriptors (DeepBit) \cite{DeepBit},
  Unsupervised Hashing with Binary Deep Neural Network (UH-BDNN) \cite{UH-BDNN},
  Similarity-Adaptive Deep Hashing (SADH) \cite{SADH},
  and GreedyHash \cite{GreedyHash}.
  Following their original papers, UH-BDNN uses the VGG feature, and the others use the raw images.
  \item Two social image hashing methods:
  Semantic-Aware Hashing (SAH) \cite{SaH} and Weakly-supervised Multimodal Hashing (WMH) \cite{WMH}.
  As SAH and WMH are shallow hashing methods, we use VGG feature as the input in our experiments to be fair.
  \end{itemize}

\subsection{Implementation Details}
$\alpha$, $\beta$ and $\nu$ balance the effects of each regularization term in the objective function.
$m$ and $a$ are the number of anchor points and concepts, respectively.
$\rho$ and $\mu$ are for ALM method.
In our experiments, the optimal performance of DSTDH is achieved
when
\{$\alpha=0.01$, $\beta=0.001$, $\nu=0.001$, $m=100$, $a=500$, $\rho=1.1$, $\mu=0.01$\}
and
\{$\alpha=10$, $\beta=0.0001$, $\nu=10$, $m=700$, $a=700$, $\rho=2$, $\mu=1$\}
on MIR Flickr and NUS-WIDE, respectively.
Our DSTDH is implemented through the open source Caffe \cite{caffe} framework.
Specifically, the mini-batch size is set as 15,
the learning rate is set as 0.01,
the momentum is set as 0.9,
and the weight decay is set as 0.0005.
We implement SAH and WMH according to their corresponding papers
since the codes of them are not available.
The parameters of the two social image hashing methods are carefully tuned to fit the datasets.
For other baseline methods,
we employ the codes provided by authors for our comparison experiments.

\subsection{Performance Comparison}
We first present the comparison results of DSTDH and its shallow variant DSTDH-L
with the baselines in the first three categories.
Then, the comparison results with social image hashing methods SAH and WMH are provided.
All the experiments are performed with 16, 32, 64 and 128 hash code length on NUS-WIDE and MIR Flickr.

\subsubsection{Comparison Results with Unsupervised Hashing}
\begin{table}
\centering
\caption{{Comparison results with social image hashing.}}
\setlength{\tabcolsep}{2.3mm}{
\begin{tabular}{c|c|c|c|c|c}
\hline
{Methods} & {Feature}& \multicolumn{4}{c}{MAP}  \\
\cline{3-6} & & 16 bits & 32 bits & 64 bits & 128 bits  \\ \hline \hline
\multicolumn{6}{c} {MIR Flickr}  \\ \hline
SAH    & VGG &	0.7360	&	0.7364	&	0.7355	&	0.7434	\\ \hline
WMH    & VGG &	0.6412	&	0.6686	&	0.6787	&	0.6951	\\ \hline
DSTDH & Raw &	\textbf{0.7664} &	\textbf{0.7769} &	\textbf{0.7893}	&	\textbf{0.7980}\\\hline
\multicolumn{6}{c} {NUS-WIDE}  \\ \hline
SAH    & VGG &	0.4398	&	0.4489	&	0.4664	&	0.4840	\\ \hline
WMH    & VGG &	0.4874	&	0.4806	&	0.4647	&	0.4427	\\ \hline
DSTDH     & Raw &\textbf{0.6324}&\textbf{0.6442}&\textbf{0.6526}&\textbf{0.6551}	\\\hline
\end{tabular}}\label{table8}
\end{table}
In Table \ref{table4} and Table \ref{table5}, we show the MAP comparison results.
In Fig. \ref{FigPR_nus}, we present the precision-recall curves of
the unsupervised shallow hashing methods (with VGG feature as input) and the unsupervised deep hashing methods.
Based on these results, we have the following observations:
(a) The MAP values of our DSTDH method consistently outperform the compared methods in all cases. The DSTDH-L with linear hash function achieves better values
than all the shallow methods with hand-crafted and VGG feature as input.
Those results clearly demonstrate the effectiveness of our methods
for hash function learning by exploiting the denoised social tags and dual-level semantic transfer.
(b) DSTDH achieves higher MAP values than DSTDH-L.
This is because DSTDH jointly performs deep image representation learning and hash function learning.
The learned image feature representation can be compatible with the hash function.
(c) The MAP values of DSTDH increase when the hash code length becomes longer,
while that of several methods (e.g. UH-BDNN with the raw image, LSMH with deep feature) gradually decrease.
These results indicate that more discriminative semantics can be obtained by our DSTDH method
if longer hash codes are adopted.
(d) Our method using the shorter code length can obtain better performance than the baseline methods using the longer code length.
For instance, on NUS-WIDE, the MAP value of DSTDH is 0.6442 when the code length is 32 bits,
which is higher than 0.6002 obtained by SADH of 128 bits.
(e) In most cases, the MAP values of the shallow methods using the VGG feature are significantly higher than those using the BOVW feature.
This is because the deep neural network has better representation capability.
(f) The precision-recall curves show that, in most situations, the area of DSTDH is larger than other competing approaches.
This indicates that our DSTDH method can return more semantically similar images to the query at the top of the retrieval results compared with the baselines.
\begin{table}
\caption{Ablation experiment results on MIR Flickr.
}
\centering
\setlength{\tabcolsep}{2mm}{
\begin{tabular}{l|c|c|c|c|c}
\hline
{Methods} & {Feature}& \multicolumn{4}{c}{MAP}  \\
\cline{3-6} & & 16 bits & 32 bits & 64 bits & 128 bits  \\ \hline \hline
DSTDH-D   & Raw &	0.7537 	&	0.7590 	&	0.7710 	&	0.7835 	\\
DSTDH-I   & Raw &	0.7541 	&	0.7751 	&	0.7799 	&	0.7939 	\\
DSTDH-NT  & Raw &	0.7499 	&	0.7583 	&	0.7712 	&	0.7843 	\\
DSTDH-T   & Raw &	0.7472 	&	0.7527 	&	0.7657 	&	0.7839 	\\
DSTDH-R   & Raw &	0.6790 	&	0.6851 	&	0.7058 	&	0.7319 	\\ \hline
DSTDH     & Raw &	\textbf{0.7664} &	\textbf{0.7769} &	\textbf{0.7893}	&	\textbf{0.7980}\\
\hline
\end{tabular}
}
\label{table6}
\end{table}

\subsubsection{Comparison Results with Social Image Hashing}
The MAP comparison results are presented in Table \ref{table8}.
It shows that the MAP values of DSTDH are higher
than those of SAH and WMH on all hash code lengths and datasets.
The superior performance is attributed to four reasons:
1) WMH and SAH directly exploit the noisy social tags during the hash learning process,
which may pose negative impact on the hashing performance.
2) With joint representation learning and hash code learning,
the hash codes generated by DSTDH
are empowered with better representation capability.
3) DSTDH adopts discrete optimization strategy to efficiently solve hash codes within a single code-solving operation,
while SAH and WMH use the relaxed optimization strategy,
which may result in significant quantization loss.
4) DSTDH jointly performs the deep hash function and code learning.
Besides, it adopts an effective dual-level semantic transfer to fully exploit the semantics of images and tags.

\subsection{Ablation Analysis}
We provide the ablation analysis on the five variant approaches.
All the experiments are conducted on MIR Flickr with different hash code lengths (16, 32, 64 and 128 bits)
and the experiment results are presented in Table \ref{table6}.
Note that similar results can be found on NUS-WIDE.
We omit them due to the space limitation.

\subsubsection{Effects of Dual-level Semantic Transfer}
To validate the effects of dual-level semantic transfer,
we compare the proposed DSTDH with its variants DSTDH-D and DSTDH-I which
remove the direct semantic transfer part and the indirect semantic transfer part, respectively.
From Table \ref{table6}, we can clearly find that DSTDH outperforms DSTDH-D and DSTDH-I on all code lengths.
These results demonstrate that the two-levels of semantic transfer are complementary to each other,
and they are effective on transferring the semantics from the social tags into the binary hash codes.

\subsubsection{Effects of Tag Guidance}
To evaluate the effects of tag guidance on hash learning,
we compare the performance with the variant approach DSTDH-NT without any tags.
The experimental results in Table \ref{table6} show that DSTDH performs better than DSTDH-NT,
which infer that using the tags (with noise removal) can improve the hashing performance.

\subsubsection{Effects of Noise Removal}
\begin{figure}
\centering
\subfigure[$\alpha$]{\includegraphics[scale=0.20]{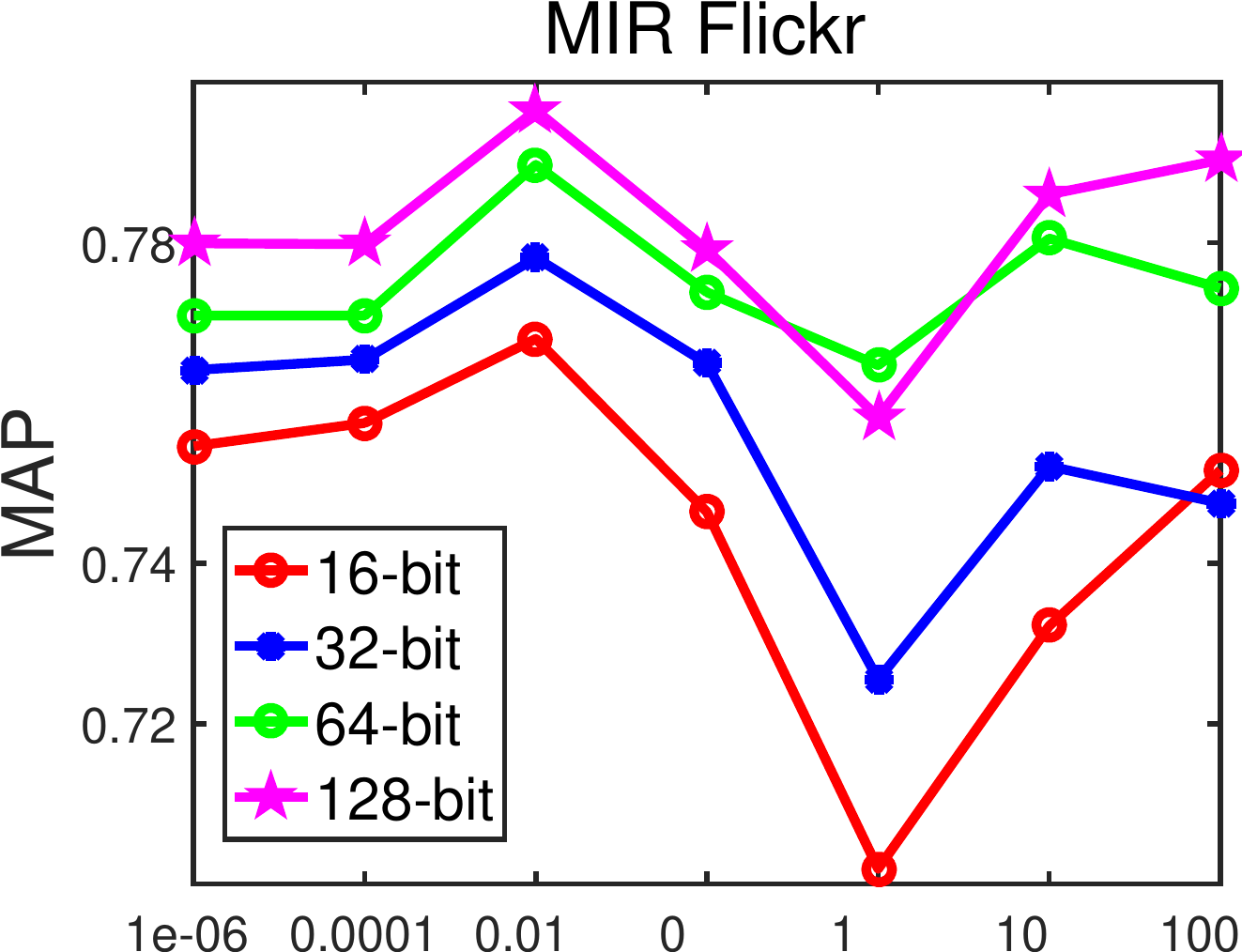}}
\subfigure[$\beta$]{\includegraphics[scale=0.20]{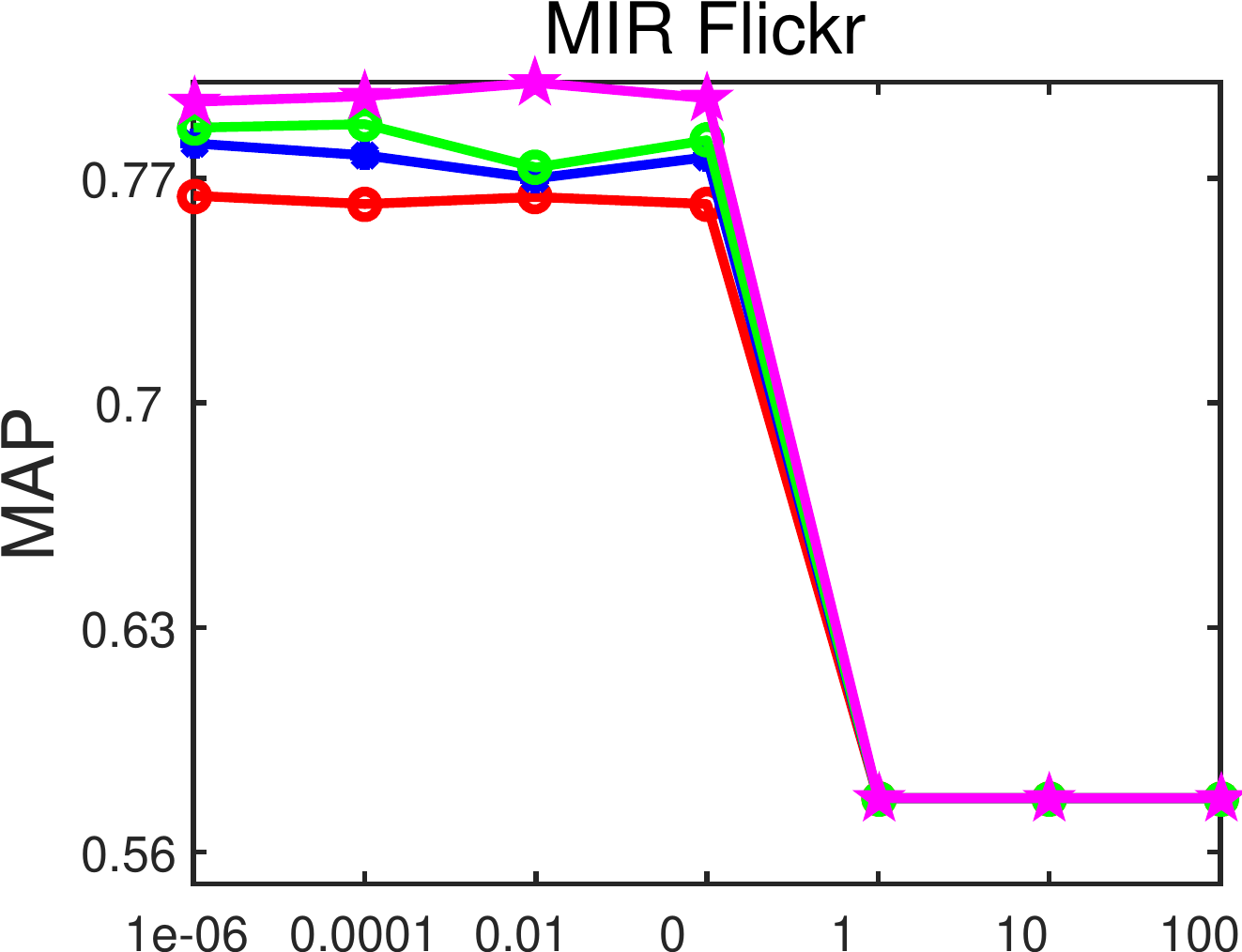}}
\subfigure[$\nu$]{\includegraphics[scale=0.20]{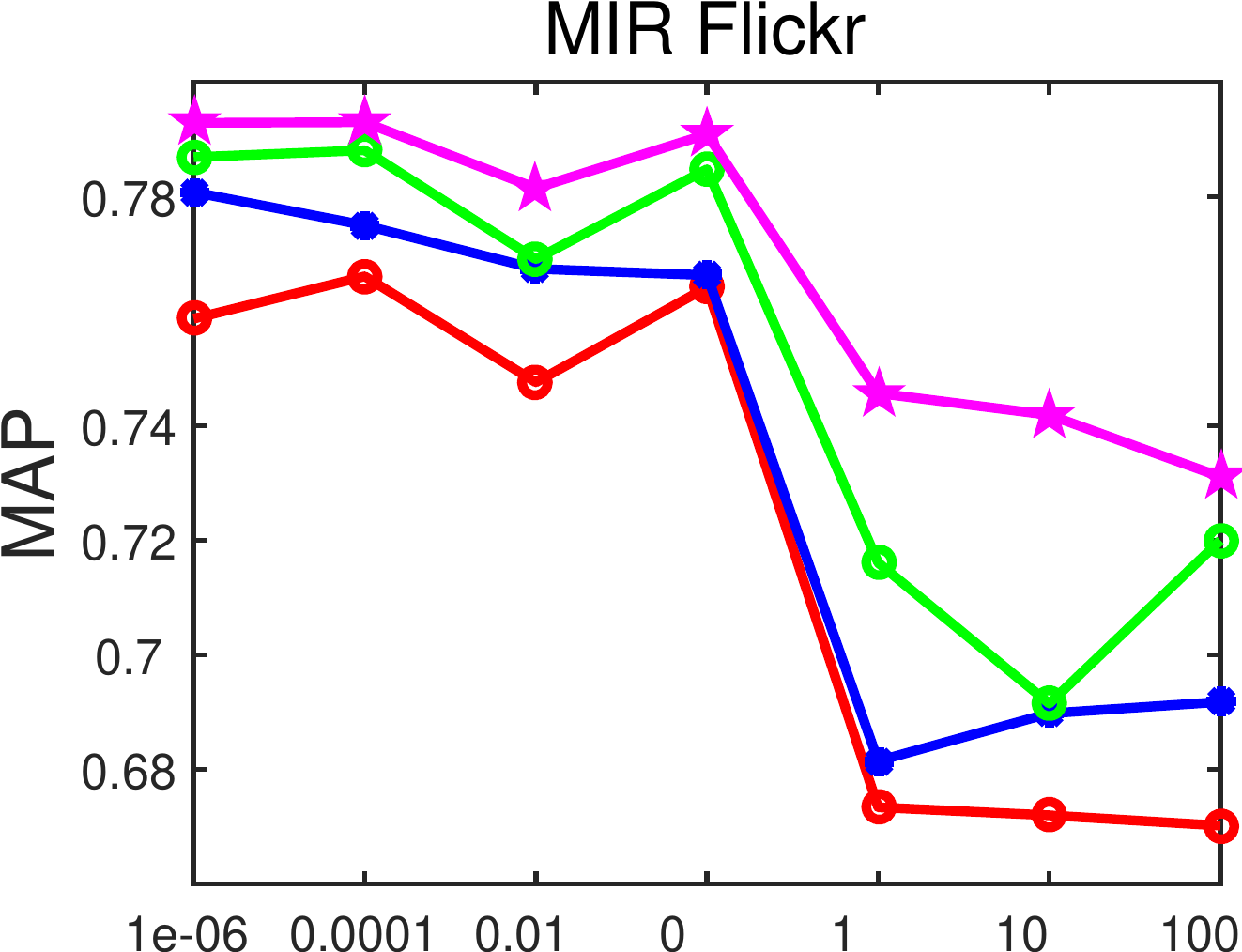}}\\
\subfigure[$m$]{\includegraphics[scale=0.25]{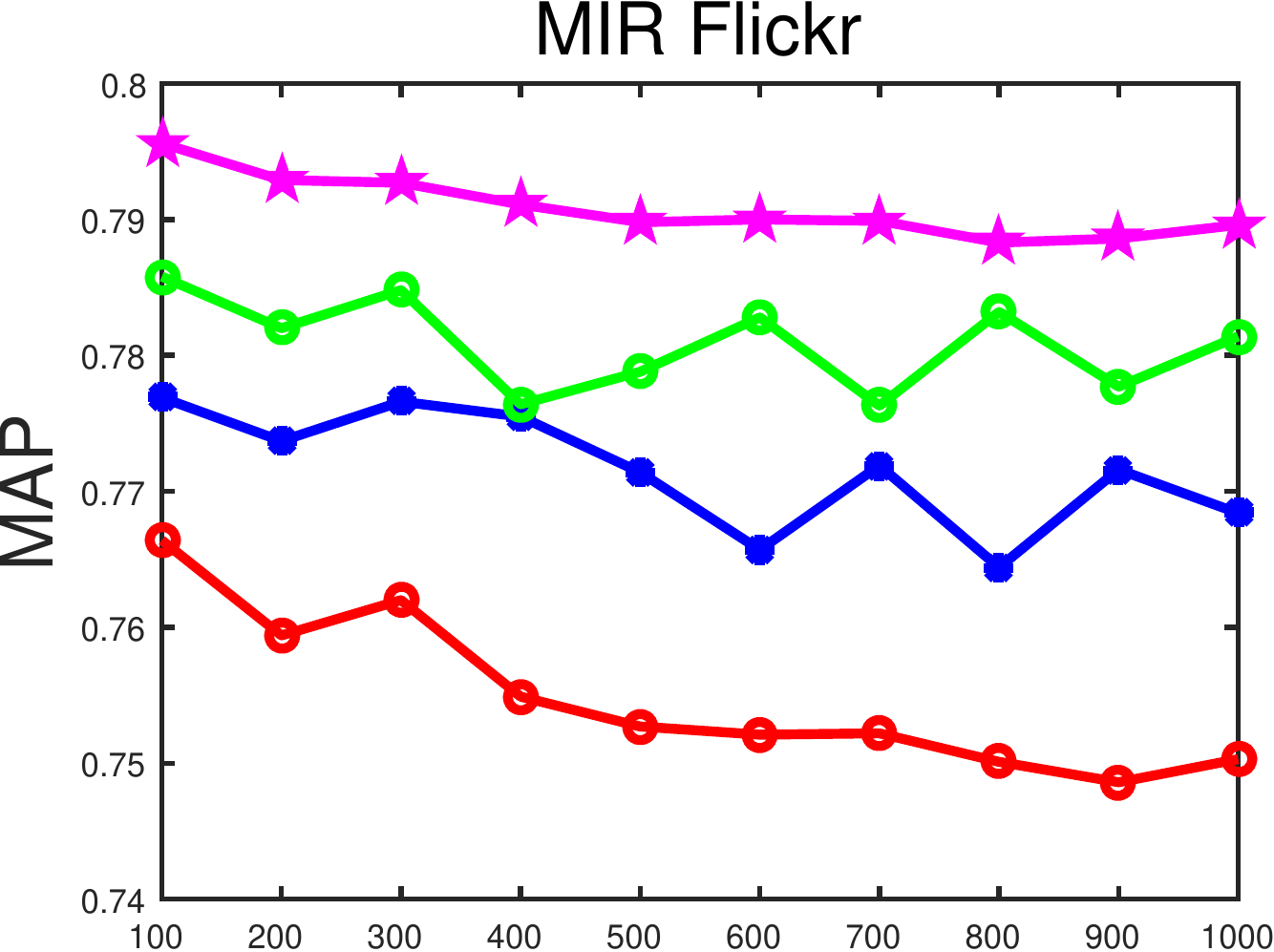}}
\subfigure[$a$]{\includegraphics[scale=0.25]{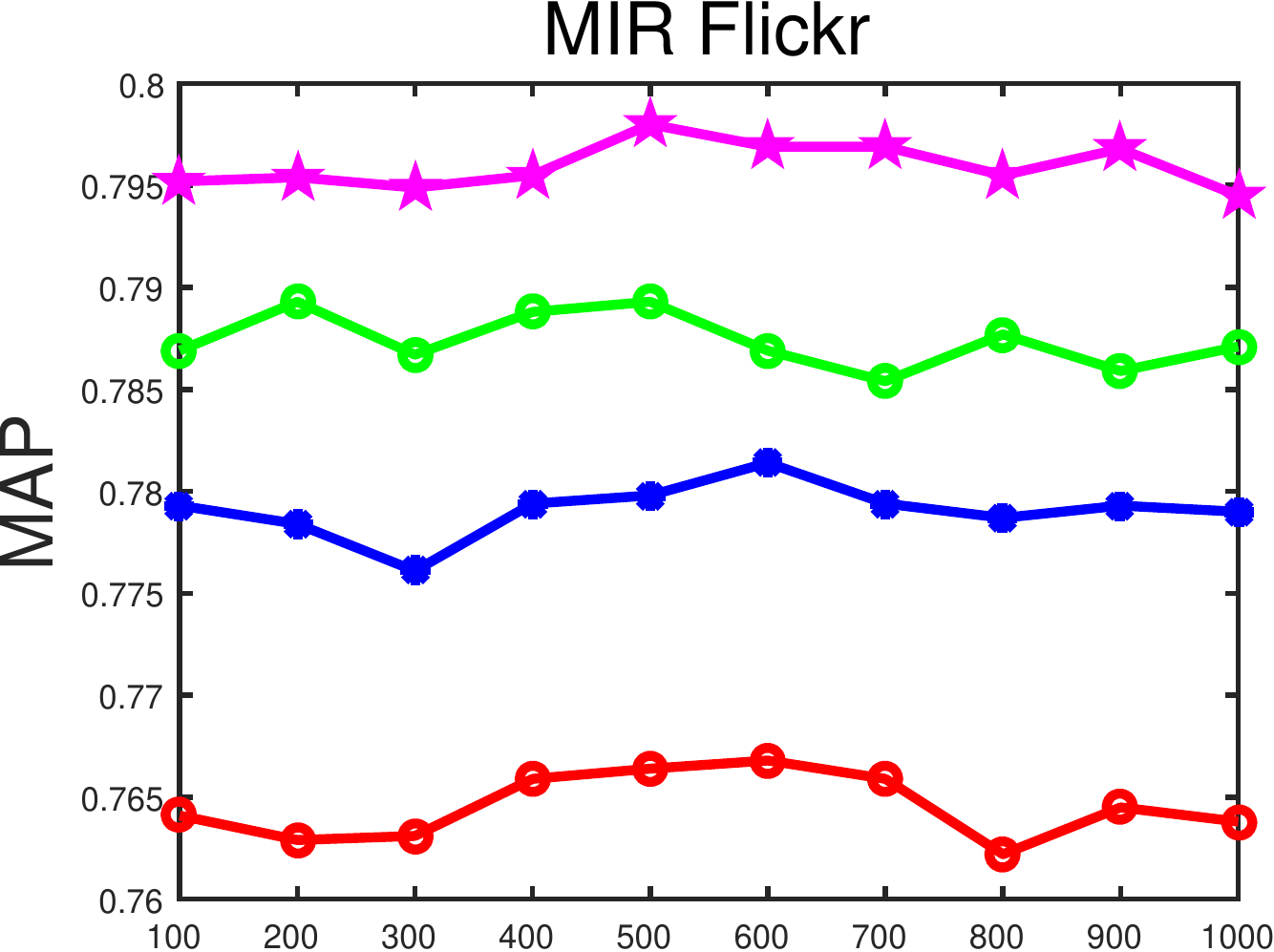}}
\caption{MAP value variations with parameters of DSTDH.}
\label{mirpara}
\end{figure}

To demonstrate the effects of $\ell _{2,1}$-norm on refining the user-generated tags,
we compare DSTDH with its variant method DSTDH-T which directly correlates original tags with hash codes without removing the noises.
As can be seen in Table \ref{table6}, DSTDH consistently outperforms DSTDH-T on different hash code lengths.
Thus, we conclude that DSTDH can effectively alleviate the negative effects of noisy semantics on the hash codes.
Besides, we can find that the MAP values of DSTDH-T are inferior to DSTDH-NT.
This phenomenon illustrates that the noises contained in user-provided tags have adverse effects on hashing performance.

\subsubsection{Effects of Discrete Optimization}
To demonstrate the effectiveness of the proposed discrete optimization strategy,
we utilize the variant approach DSTDH-R for comparison.
It firstly relaxes the discrete constraints on hash codes, and then obtains the final hash codes by mean-thresholding.
In Table \ref{table6}, the comparison results show that DSTDH is obviously superior to DSTDH-R.
This phenomenon reveals that discrete optimization can effectively
alleviate the binary quantization loss and improve the retrieval performance.

\subsection{convergence and Parameter Analysis}
In this subsection, we first analyze the convergence of our DSTDH method
on MIR Flickr when the hash code length is fixed as 32 bits.
The convergence curve is shown in Fig. \ref{mirpara}(a).
It can be seen that the objective function value first decreases monotonically,
and eventually reaches the local minimum after several iterations (less than 10).
These results verify the convergence of our proposed discrete hash optimization strategy.

Then, we conduct parameter sensitivity analysis experiments to observe the variation of MAP values
under different $\alpha$, $\beta$, $\nu$, {$m$ and $a$}.
The experiments are performed on MIR Flickr when the hash code length varying from 16 to 128 bits.
We vary the value of $\alpha$, $\beta$, $\nu$ from $\{10^{-6}, 10^{-4}, 10^{-2}, {0}, 10^{0}, 10^1, 10^2\}$,
while the others are fixed.
{Moreover, the number of anchor $m$ and concept $a$ are both varied from the range of $\{100, 200, ..., 1000\}$ by
fixing other parameters.}

From Fig. \ref{mirpara} (a) - (c), we can find that
when $\alpha$ is in the range of $\{10^{-6}, {0}\}$,
$\beta$ is in the range of $\{10^{-6}, {0}\}$
and $\nu$ is in the range of $\{10^{-6}, {0}\}$,
the performance is relatively stable.
{From Fig. \ref{mirpara} (d), we can find that the MAP values fluctuate in several hash code lengths but the overall trend is decreasing with the number of anchor points increasing.
From Fig. \ref{mirpara} (e), we can find that the MAP values are relatively stable when $a$ is varied from 100 to 1000 on all hash code lengths.
}


\subsection{Time Cost with Training Size}
\begin{figure}
\centering
\mbox{
\subfigure[]{\includegraphics[scale=0.32]{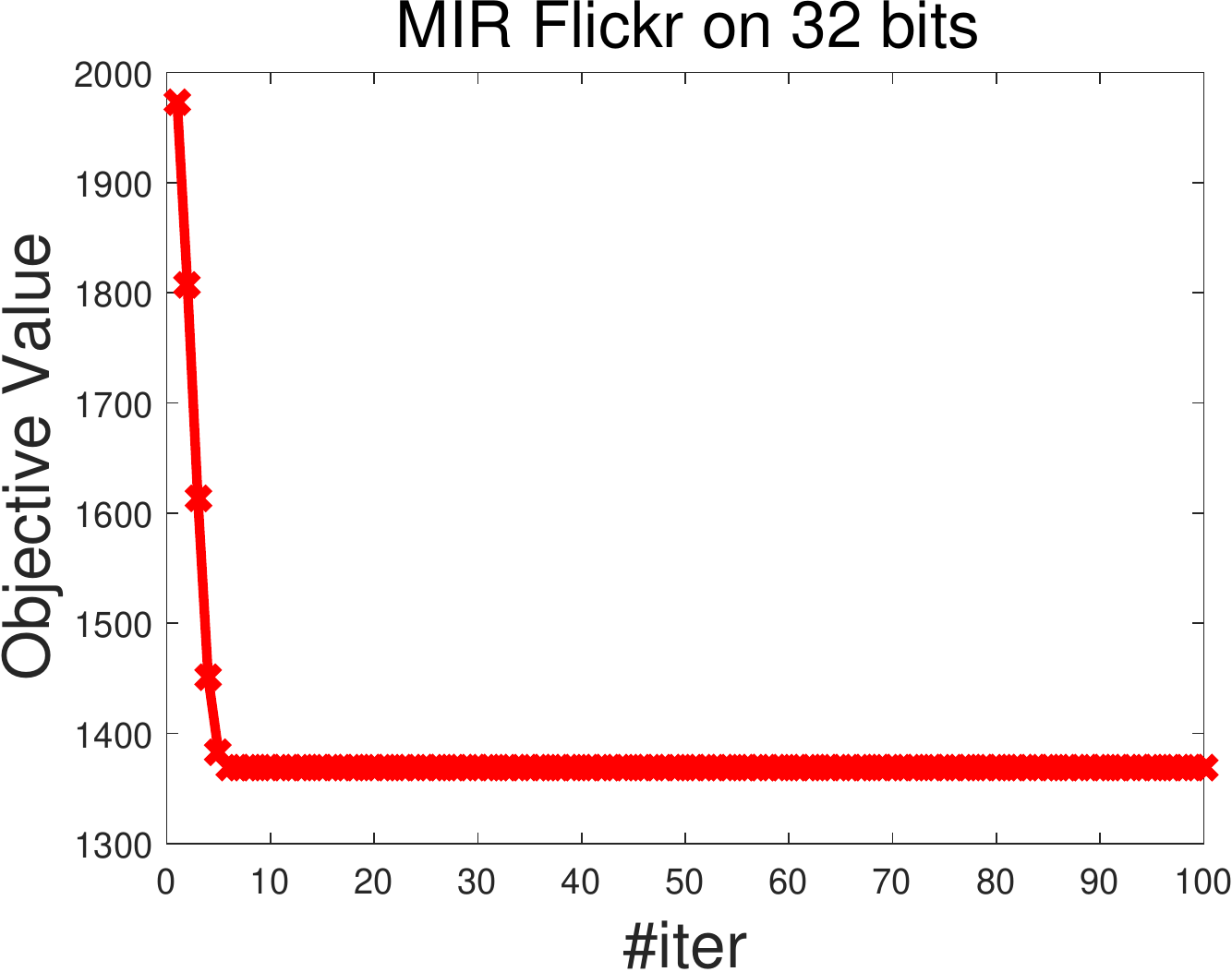}}
}\mbox{
\subfigure[]{\includegraphics[scale=0.32]{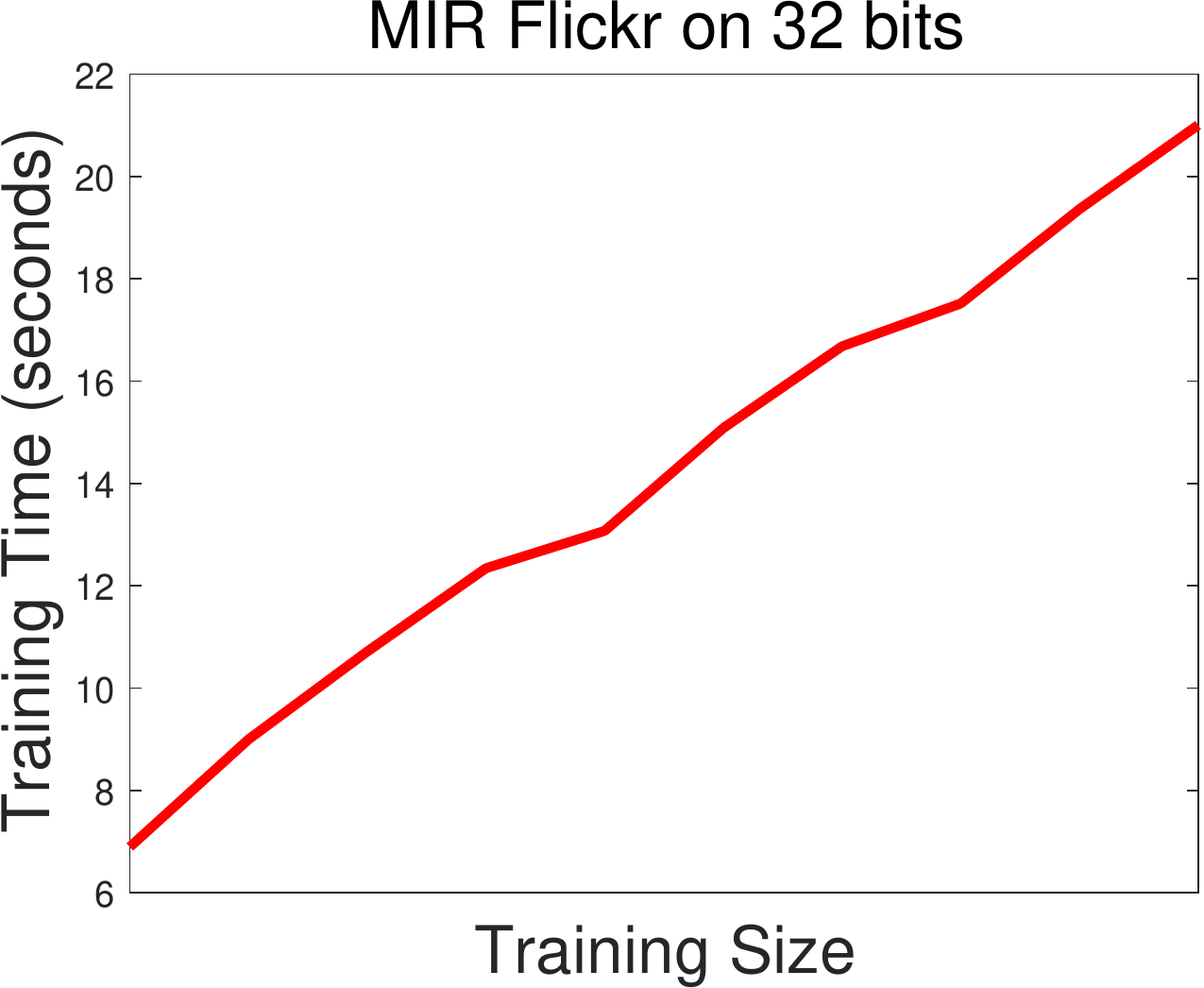}}
}
\caption{(a) Objective function value variations with the number of iterations.
(b) Training time cost variations with the training size.}
\label{size_conv}
\end{figure}

In this subsection, we investigate the training time cost variations with the training size.
Specifically, we conduct experiment on MIR Flickr and record time cost
when the training size is varied from 500 to 5,000
by fixing the hash code length as 32.
Fig. \ref{size_conv} (b) demonstrates the training time increases linearly with the training size.
It validates the linear scalability of DSTDH,
which is consistent with the aforementioned theoretical analysis.


\section{Conclusion} \label{conc}
In this paper, we propose a \emph{Dual-level Semantic Transfer Deep Hashing} method
for efficient social image retrieval.
We develop a weakly-supervised deep hash learning framework that simultaneously learns powerful image representation and discriminative hash codes
by exploiting the social tags (with noise removal) and considering dual-level semantic transfer.
Therefore, the semantics of tag can be efficiently discovered and seamlessly transferred into the binary hash codes.
Moreover, we directly solve the binary hash codes by a discrete optimization strategy
to alleviate the relaxing quantization errors and thus guarantee the transfer effectiveness.
Extensive experiments on two public social image retrieval benchmarks demonstrate the superiority of the proposed method.

In this work, we adopt $\ell _{2,1}$-norm to remove the noises from social tags,
directly pack the visual feature matrix and the image-tag relation matrix
for hypergraph construction, and choose simple Euclidean loss to optimize the model.
In the further, we will explore more effective denoising method,
hypergraph construction method, and loss function into our framework to improve the performance further.




\bibliographystyle{IEEEtran}
\bibliography{sample-base}
\newpage
\begin{IEEEbiography}[{\includegraphics[width=1in,height=1.25in,clip,keepaspectratio]{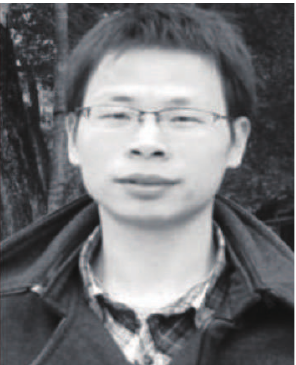}}]{Lei Zhu}
is currently a professor with the School of Information Science and Engineering, Shandong Normal University. He received his B.Eng. and Ph.D. degree from Wuhan University of Technology in 2009 and Huazhong University Science and Technology in 2015, respectively. He was a Research Fellow at the Singapore Management University (2015-2016), and at the University of Queensland (2016-2017). His research interests are in the area of large-scale multimedia content analysis and retrieval. Dr. Zhu has co-/authored more than 80 peer-reviewed papers, like CVPR, ICCV, ACM MM, SIGIR, TPAMI, TIP, TMM, TCSVT. He is an Associate Editor of Neural Processing Letters, and The Journal of Electronic Imaging.  He has served as the PC member
for several top conferences such as MM, SIGIR, etc., and the regular reviewer
for journals including TIP, TCSVT, TMM, etc. He was granted several awards like SIGIR 2019 best paper honorable mention.
\end{IEEEbiography}
\begin{IEEEbiography}[{\includegraphics[width=1.2in,height=1.25in,clip,keepaspectratio]{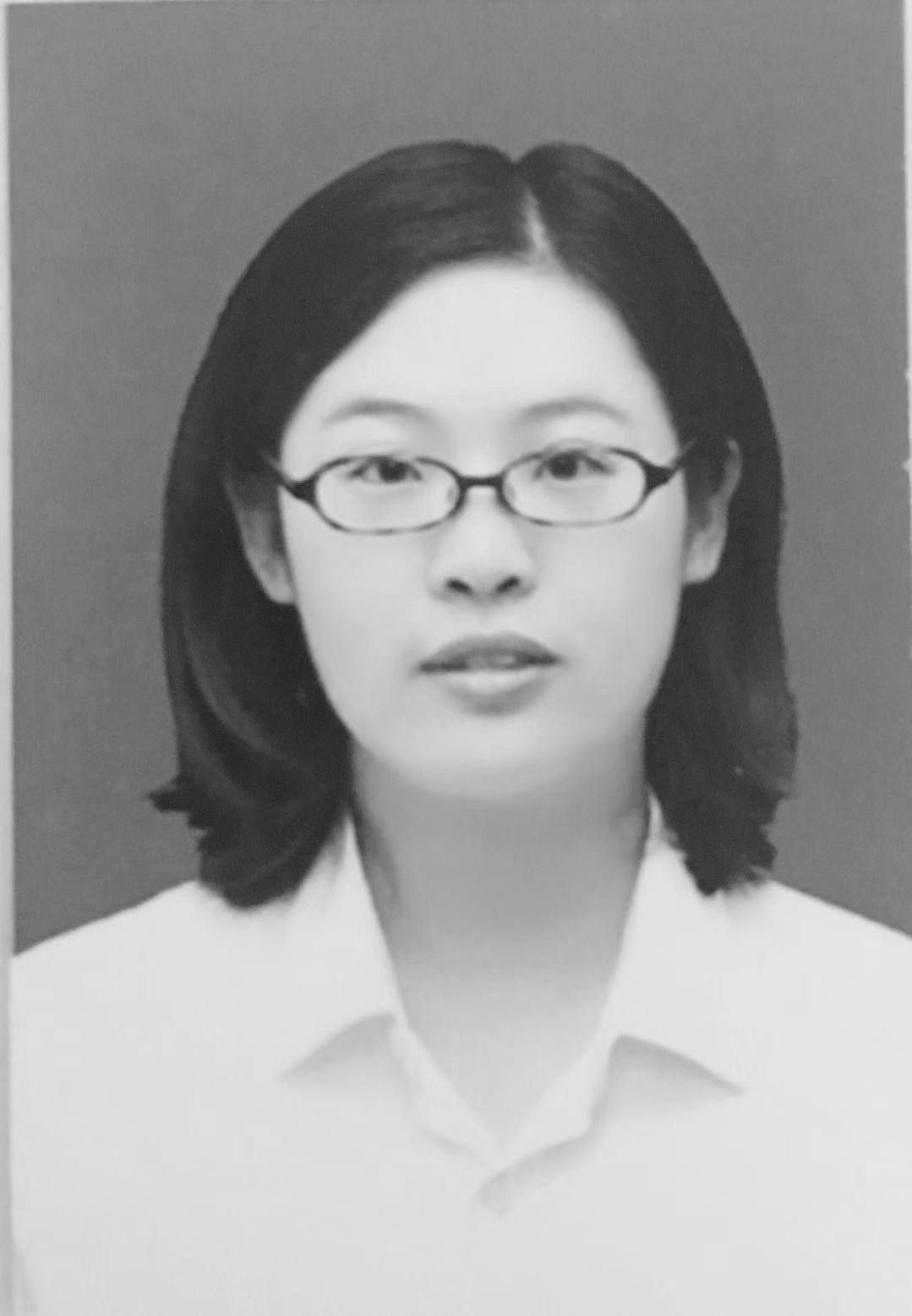}}]{Hui Cui}
is with the School of Information Science and Engineering, Shandong Normal University, China. Her research interest is in the area of large-scale multimedia content analysis and retrieval.
\end{IEEEbiography}


\begin{IEEEbiography}[{\includegraphics[width=1.2in,height=1.25in,clip,keepaspectratio]{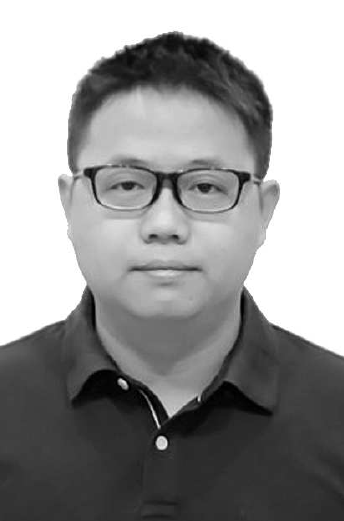}}]{Zhiyong Cheng}
is currently a Professor with Shandong Artificial Intelligence Institute, Qilu University of Technology (Shandong Academy of Sciences). He received the Ph.D degree in computer science from Singapore Management University in 2016, and then worked as a Research Fellow in National University of Singapore. His research interests mainly focus on large-scale multimedia content analysis and retrieval. His work has been published in a set of top forums, including ACM SIGIR, MM, WWW, TOIS, IJCAI, TKDE, and TCYB. He has served as the PC member for several top conferences such as MM, MMM etc., and the regular reviewer for journals including TKDE, TIP, TMM etc.
\end{IEEEbiography}
\vfill
\vspace{-4mm}
\newpage
\begin{IEEEbiography}[{\includegraphics[width=1.2in,height=1.25in,clip,keepaspectratio]{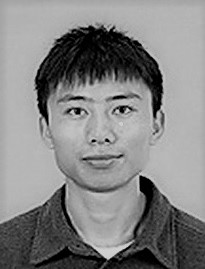}}]{Jingjing Li}
received his MSc and PhD degree in Computer Science from University of Electronic Science and Technology of China in 2013 and 2017, respectively. Now he is a national Postdoctoral Program for Innovative Talents research fellow with the School of Computer Science and Engineering, University of Electronic Science and Technology of China. He has great interest in machine learning, especially transfer learning, subspace learning and recommender systems.
\end{IEEEbiography}

\begin{IEEEbiography}[{\includegraphics[width=1.2in,height=1.25in,clip,keepaspectratio]{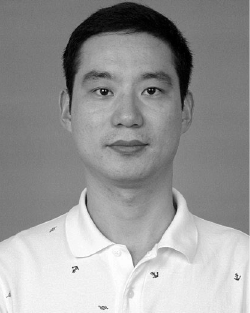}}]{Zheng Zhang}
received his M.S. degree in Computer Science (2014) and Ph.D. degree in Computer Applied Technology (2018) from the Harbin Institute of Technology, China. Dr. Zhang was a Postdoctoral Research Fellow at The University of Queensland, Australia. He is currently an Assistant Professor at Harbin Institute of Technology, Shenzhen, China. He has published over 50 technical papers at prestigious international journals and conferences, including the IEEE TPAMI, IEEE TNNLS, IEEE TIP, IEEE TCYB, IEEE CVPR, ECCV, AAAI, IJCAI, SIGIR, ACMM, etc. He serves/served as a (leading) Guest Editor of Information Processing and Management journal and Neurocomputing journal, a Publication Chair of the 16th International Conference on Advanced Data Mining and Applications (ADMA 2020), and an SPC/PC member of several top conferences. His current research interests include machine learning, computer vision and multimedia analytics.\end{IEEEbiography}
\vfill

\end{document}